\documentclass[10pt,preprint,eqsecnum]{aastex}
\usepackage{amsmath}
\begin{document}
%------------------------------------------------------------------------------
%\markright{Instability of ...}
%------------------------------------------------------------------------------

\title{Non-Oscillatory Central Difference and
Artificial Viscosity Schemes for Relativistic Hydrodynamics}

\author{Peter Anninos and P. Chris Fragile}
\affil{University of California,Lawrence Livermore National Laboratory,
Livermore CA 94550 }

%\date{{\small    \today}}
%\date{{\small   \LaTeX-ed \today}}
%-----------------------------------------------------------------------------

\begin{abstract}
High resolution, non-oscillatory, central difference 
(NOCD) numerical schemes
are introduced as alternatives to more traditional artificial
viscosity (AV) and Godunov methods for solving
the fully general relativistic hydrodynamics equations.
These new approaches provide the advantages of Godunov
methods in capturing ultra-relativistic flows without the
cost and complication of Riemann solvers, and the advantages
of AV methods in their speed, ease of
implementation, and general applicability without
explicitly using artificial viscosity for shock capturing. Shock tube,
wall shock, and dust accretion tests, all with adiabatic
equations of state, are presented
and compared against equivalent solutions from
both AV and Godunov based codes.
In the process we address the accuracy
of time-explicit NOCD and AV methods over a wide range of Lorentz factors.
\end{abstract}


\section{Introduction}
\label{sec:intro}

The earliest attempts at simulating relativistic flows
in the presence of strong gravitational fields are attributed to
May and White (1966, 1967) who investigated gravitational
collapse in a one dimensional Lagrangian code using 
artificial viscosity (AV) methods \citep{Neumann50} to capture
shock waves.
Wilson (1972, 1979) subsequently
introduced an alternative Eulerian coordinate
approach in multi-dimensional calculations, using
traditional finite difference upwind methods and 
artificial viscosity for shock capturing.
Since these earliest studies, AV methods have continued 
to be developed in their popularity and applied to 
a variety of problems due
largely to their general robustness
\citep{HSW84_1,HSW84_2,CW84,Anninos98}.
These methods are also computationally cheap, 
easy to implement, and easily adaptable to multi-physics applications.
However, it has been demonstrated that problems
involving high Lorentz factors (greater than a few)
are particularly sensitive to different implementations of
the viscosity terms, and can result in large numerical errors
if solved using time explicit methods \citep{Norman86}.

Significant progress has been made in recent years to take
advantage of the conservational form of the hydrodynamics system
of equations to apply Godunov-type methods 
and approximate Riemann solvers to simulate
ultra-relativistic flows \citep{Eulderink95,Banyuls97,Font00}.
Although Godunov-based schemes are accepted as more
accurate alternatives to AV methods, especially in the limit
of high Lorentz factors, they are not
infallible and should generally be used with caution.
They may produce unexpected
results in certain cases that can be overcome only
with specialized fixes or by adding additional dissipation.
A few known examples include the admittance of expansion shocks,
negative internal energies in kinematically dominated flows,
`carbuncle' effect in high Mach number bow shocks,
kinked Mach stems, and odd/even decoupling in
mesh-aligned shocks \citep{Quirk94}.
Godunov methods, whether they solve the Riemann problem exactly
or approximately, are also computationally much more expensive than
their simpler AV counterparts, and more difficult to extend the
system of equations to include additional physics.

Hence we have undertaken this current study to explore an
alternative approach of using high resolution, non-oscillatory,
central difference (NOCD) methods \citep{Jiang98a,Jiang98b}
to solve the relativistic hydrodynamics
equations. These new schemes combine the speed, efficiency, and
flexibility of AV methods
with the advantages of the potentially more
accurate conservative formulation approach of Godunov methods,
but without the cost and complication of Riemann solvers
and flux splitting.

The NOCD methods are implemented as part of a new code
we developed called Cosmos, and designed for fully general
relativistic problems. 
Cosmos is a collection of massively
parallel, multi-dimensional, multi-physics solvers applicable
to both Newtonian and general relativistic systems, 
and currently includes
five different computational fluid dynamics (CFD) methods, equilibrium
and non-equilibrium
primordial chemistry, photoionization, radiative cooling,
radiation flux-limited diffusion, radiation pressure,
scalar fields, Newtonian external and self gravity, 
arbitrary spacetime geometries,
and viscous stress. The five hydrodynamics
methods include a Godunov (TVD) solver for Newtonian flows,
two artificial viscosity codes for general relativistic
systems (differentiated by mesh
or variable centering type: staggered versus zone-centered),
and two relativistic methods based on non-oscillatory
central difference schemes (differentiated also by the
mesh type: staggered versus centered in time and space).
The emphasis in the following sections is to review our
particular implementations of the AV and NOCD methods
and compare results of various shock wave and
accretion test calculations with other published results.
We also explore the accuracy of
both AV and NOCD methods in simulating
ultra-relativistic shocks over a wide range of Lorentz factors.

\section{Basic Equations}
\label{sec:equations}

\subsection{Internal Energy Formulation}
\label{sec:internal_e}

Both of the artificial viscosity methods in Cosmos are based on an
internal energy formulation of the perfect fluid conservation equations.
The equations are derived
from the 4-velocity normalization $u^{\mu}u_{\mu} = -1$,
the conservation of baryon number $\nabla_{\mu} (\rho u^{\mu}) = 0$
for the fluid rest mass density,
the parallel component of the stress--energy conservation equation
$u_\nu\nabla_{\mu} T^{\mu\nu} = 0$ for internal energy,
the transverse component of the stress--energy conservation equation
$(g_{\alpha\nu} + u_\alpha u_\nu) \nabla_{\mu} T^{\mu\nu} = 0$ for momentum,
and an adiabatic equation of state (eos)
for the fluid pressure 
$P=(\Gamma-1) e$, 
where $\Gamma$ is the adiabatic index and
$e$ is the fluid internal energy density.
For a perfect fluid, the stress-energy tensor is
\begin{equation}
T^{\mu\nu} = \rho h u^\mu u^\nu + P g^{\mu\nu} ,
\label{eqn:tmn}
\end{equation}
where 
\begin{equation}
h = 1 + \epsilon + \frac{P}{\rho} = 1 + \Gamma\epsilon
\end{equation}
is the relativistic enthalpy,
$\epsilon$ is the specific internal energy, $u^\mu$ is the contravariant
4-velocity, and $g_{\mu\nu}$ is the 4-metric.
The resulting equations can be written in flux conservative form as
\citep{Wilson79}
\begin{eqnarray}
 \frac{\partial D}{\partial t} + \frac{\partial (DV^i)}{\partial x^i} &=& 0 ,
      \label{eqn:av_de} \\
 \frac{\partial E}{\partial t} + \frac{\partial (EV^i)}{\partial x^i} 
 + P\frac{\partial W}{\partial t} + P\frac{\partial (WV^i)}{\partial x^i} &=& 0
      \label{eqn:av_en} \\
 \frac{\partial S_j}{\partial t} + \frac{\partial (S_j V^i)}{\partial x^i}
 - \frac{S^\mu S^\nu}{2S^0} \frac{\partial g_{\mu\nu}}{\partial x^j} 
 + \sqrt{-g} \frac{\partial P}{\partial x^j} &=& 0 ,
      \label{eqn:av_mom}
\end{eqnarray}
where $g$ is the determinant of the 4-metric,
$W=\sqrt{-g} u^0$ is the relativistic boost factor, 
$D=W\rho$ is the generalized fluid density, 
$V^i=u^i/u^0$ is the transport velocity,
$S_i = W\rho h u_i$ is the covariant momentum density, 
and $E=We=W\rho\epsilon$ is the generalized internal energy density.
We use the standard convention in which Greek (Latin) 
indices refer to 4(3)-space components.

The system of equations (\ref{eqn:av_de}) -- (\ref{eqn:av_mom}) 
are complemented by two additional expressions
for $V^i$ and $W$ that are convenient for numerical computation.
Introducing a general tensor form for artificial viscosity $Q_{ij}$ 
(see section \ref{sec:artificial}), and defining
\begin{equation}
M = \rho h W = \rho h \sqrt{-g} u^0 = E + D + (P+\mbox{tr}[Q_{ij}]) W \space ,
\label{eqn:m2}
\end{equation}
the momentum can be expressed as $S_\mu = M u_\mu$, and
$S_0$ is computed
from the normalization of the four--velocity $S^\mu S_\mu = -M^2$.
The coordinate velocity then becomes $V^i= S^i/S^0$ with $V^0=1$.
Also, the time component of the four--velocity
$u^0$ can be calculated from the normalization
$u_\mu u^\mu = u^0 V^\mu S_\mu/M = -1$, and used to derive
the following expressions for $W$
\begin{equation}
W = \frac{-\sqrt{-g}~M}{S_\mu V^\mu} = \frac{\sqrt{-g} S^0}{\rho h W} .
\label{eqn:w2}
\end{equation}
The former expression ($W=-\sqrt{-g}~M/(S_\mu V^\mu)$)
is used in the staggered mesh AV schemes as it results
in more accurate density and velocity jump conditions
across shock fronts. The latter is more convenient
for the zone centered NOCD methods.

\subsection{Conservative Energy Formulation}
\label{sec:conservative_e}

The second class of numerical methods presented in this paper 
(the NOCD schemes) are based
on a simpler conservative hyperbolic 
formulation of the hydrodynamics equations. In this case,
the equations are derived directly from the conservation of stress-energy
\begin{equation}
\nabla_{\mu} T^{\mu\nu} = 
   \frac{1}{\sqrt{-g}}\left(\sqrt{-g} T^{\mu\nu}\right)_{,\mu}
  +\Gamma^\nu_{\alpha\mu} T^{\mu\alpha} = 0.
\label{eqn:tmnu2}
\end{equation}
Expanding (\ref{eqn:tmnu2}) into time and space explicit parts
yields the flux conservative equations for general stress-energy
tensors
\begin{equation}
\frac{\partial (\sqrt{-g}~T^{0\nu})}{\partial t} +
\frac{\partial (\sqrt{-g}~T^{i\nu})}{\partial x^i} = \Sigma^\nu ,
\label{eqn:tmnu3}
\end{equation}
with curvature source terms
\begin{equation}
\Sigma^\nu = -\sqrt{-g}~T^{\beta\gamma}~\Gamma^\nu_{\beta\gamma}.
\end{equation}
Substituting the perfect fluid stress tensor (\ref{eqn:tmn}) 
into (\ref{eqn:tmnu3}), and including baryon conservation results
in the following set of equations
\begin{eqnarray}
 \frac{\partial D}{\partial t} + \frac{\partial (DV^i)}{\partial x^i} 
 &=& 0 ,
      \label{eqn:hr_de} \\
 \frac{\partial {\cal E}}{\partial t} 
 + \frac{\partial ({\cal E}V^i)}{\partial x^i} 
 + \frac{\partial [\sqrt{-g}~(g^{0i} - g^{00} V^i)~P]}{\partial x^i} 
 &=& \Sigma^0,
      \label{eqn:hr_en} \\
 \frac{\partial {\cal S}^j}{\partial t} 
 + \frac{\partial ({\cal S}^j V^i)}{\partial x^i} 
 + \frac{\partial [\sqrt{-g}~(g^{ij} - g^{0j} V^i)~P]}{\partial x^i} 
 &=& \Sigma^j,
      \label{eqn:hr_mom}
\end{eqnarray}
where the variables $D$, $V^i$, and $g$ are the same
as those defined in the internal energy formulation.
However, now 
\begin{eqnarray}
{\cal E}   &=& W \rho h u^0 + \sqrt{-g}~g^{00} P , \\
{\cal S}^i &=& W \rho h u^i + \sqrt{-g}~g^{0i} P ,
\end{eqnarray}
are the new expressions for energy and momenta.

It is convenient to express ${\cal E}$ and ${\cal S}^i$ in terms
of the internal energy formulation variables, especially for
initializing data 
\begin{eqnarray}
{\cal E}   &=& \frac{W^2}{\sqrt{-g}}\left(\frac{D}{W}
               + \Gamma\frac{E}{W}\right)
               + (\Gamma-1)\sqrt{-g}~g^{00} \frac{E}{W} , \\
{\cal S}^i &=& g^{i\alpha} S_\alpha + (\Gamma-1)~\sqrt{-g}~g^{0i} \frac{E}{W} ,
\end{eqnarray}
and reconstructing the equation of state
\begin{eqnarray}
P &=& (\Gamma-1)\frac{E}{W}  \\
  &=& \left(\frac{{\cal E}\sqrt{-g}}{W^2} - \frac{D}{W}\right)
      \frac{\Gamma-1}{\Gamma + (\Gamma-1) g^{00}(\sqrt{-g}/W)^2} ,
\label{eqn:eos}
\end{eqnarray}
where we have explicitly assumed an adiabatic gamma-law fluid.

\section{Numerical Methods}
\label{sec:methods}

Cosmos is a multi-dimensional (1, 2 or 3D) code that uses
regularly spaced Cartesian meshes for spatial finite differencing 
or finite volume discretization methods.
Evolved variables are defined at the
zone centers in the NOCD, TVD, and non-staggered AV methods.
In the staggered mesh AV method, variables are centered
either at zone faces (the velocity $V^j$ and momentum $S_j$ vectors)
or zone centers (all other scalar and tensor variables).
Periodic, reflection, constant in time, user-specified, and flat 
(vanishing first derivative) boundary
conditions are supported for all variables in the evolutions.
The hydrodynamic equations in both of the formalisms
presented in \S\ref{sec:equations}
are solved with time-explicit, operator split methods
with second order spatial finite differencing.
Single-step time integration and dimensional splitting is used 
for both AV methods. The NOCD schemes use a second order
predictor-corrector time integration with dimensional splitting,
and the TVD approach utilizes a third order Runge-Kutta
time integration with finite volume representations for source updates.
Since the main emphasis here is on relativistic hydrodynamics,
the following discussion is limited to presenting details relevant
for the AV and NOCD schemes: the TVD method is currently only Newtonian.

\subsection{Artificial Viscosity}
\label{sec:artificial}

The order and frequency in which various source terms 
and state variables are updated in the AV methods
can affect the numerical accuracy, especially in high boost flows.
The following order composing a complete single cycle
or time-step solution has been determined to produce a reasonable
compromise between cost and accuracy:
\begin{itemize}
\item[$\bullet$]
  Compute time step $\Delta t$ from (\ref{eqn:timestep})
\item[$\bullet$]
  Store current value of boost factor $W$
\item[$\bullet$]
  Curvature
     \begin{itemize}
     \item[\bf --]
	compute pressure and sound speed from the ideal 
	fluid equation of state: \\
	$P=(\Gamma-1)E/W$, $C_s = \sqrt{\Gamma P/(\rho h)}$
     \item[\bf --]
	evaluate scalar or tensor artificial viscosity $Q_{ij}$
     \item[\bf --]
        normalize velocity and update boost factor: \\
	$V^i=S^i/S^0$, using 
	$S_\mu S^\mu = -M^2 = -(D+E+PW+\mbox{tr}[Q_{ij}]W)^2$
	to first compute $S_0$; \\
	then construct $S^\mu$ from $g^{\mu\nu}$, $S_0$, and 
	the evolved $S_j$; \\
	and finally use equation (\ref{eqn:w2}) to define the boost factor 
	$W=-\sqrt{-g}M/(S_\mu V^\mu)$
     \item[\bf --]
        update momentum, accounting for curvature: \\
        $\dot S_j = S^\mu S^\nu g_{\mu\nu,j}/(2S^0)$,
	using second order finite differencing of $g_{\mu\nu}$
     \end{itemize}
\item[$\bullet$]
  Artificial viscosity
     \begin{itemize}
     \item[\bf --]
	compute pressure
     \item[\bf --]
        normalize velocity, update $W$
     \item[\bf --]
	compute pressure and sound speed
     \item[\bf --]
	evaluate artificial viscosity components $Q_{ij}$
     \item[\bf --]
        update momentum and energy equations accounting for $Q_{ij}$: \\
        $\dot E   = -\sum_{i,j} Q_{ij} [\nabla_i (WV^j) + \nabla_j (WV^i)]/2$,
        and $\dot S_j = -\sqrt{-g} \nabla_i Q_{ij}$
     \end{itemize}
\item[$\bullet$]
  Compression
     \begin{itemize}
     \item[\bf --]
	compute pressure
     \item[\bf --]
        normalize velocity, update $W$
     \item[\bf --]
	compute pressure again
     \item[\bf --]
        update energy, accounting for compressional heating: \\
	$\dot E = -P\nabla_i(W V^i)$
     \end{itemize}

\item[$\bullet$]
  Pressure gradient
     \begin{itemize}
     \item[\bf --]
	compute pressure
     \item[\bf --]
        update momentum, accounting for pressure gradients: \\
	$\dot S_j = -\sqrt{-g} \nabla_j P$
     \end{itemize}

\item[$\bullet$]
  Transport
     \begin{itemize}
     \item[\bf --]
	compute pressure 
     \item[\bf --]
        normalize velocity, update $W$
     \item[\bf --]
	update transport terms in all variables:  \\
	$\dot D = -\nabla_i(DV^i)$, $\dot E = -\nabla_i(EV^i)$,
        and $\dot S_j = -\nabla_i(S_jV^i)$
     \end{itemize}

\item[$\bullet$]
  Boost factor
     \begin{itemize}
     \item[\bf --]
	compute pressure and sound speed
     \item[\bf --]
        normalize velocity, final update of $W$
     \item[\bf --]
        update energy, accounting for the variation of $W$ in time: \\
	$\dot E = -[P + (\sum_i Q_{ii}^2/\sum_i Q_{ii}) ] \dot W$
     \end{itemize}
\item[$\bullet$]
  Update spacetime metric components $g_{\mu\nu}$ and
  $g^{\mu\nu}$ if time dependent
\end{itemize}

The highly nonlinear coupling of pressure and artificial viscosity
to the state and kinematic variables through the Lorentz factor
makes the relativistic equations much more difficult to solve
than their Newtonian versions. It is for this reason that 
\citet{Norman86} adopted implicit methods to
solve the special relativistic equations. It is also
why we have attempted to maintain a consistent and frequent
update of the velocity normalization,
boost factor, pressure and artificial viscosity throughout the cycle.

To enforce stable evolutions,
the time step is defined for all hydro methods as the minimum
causality constraint over the entire mesh
arising from the sound speed,
fluid velocity, magnitude of the artificial viscosity coefficient,
and any other physics criteria
introduced in the calculation, say from
scalar fields, radiation transport, gravity, etc...
Also, since the time steps can be nonuniform,
a final constraint is added to prevent $\Delta t$ from
increasing by more than 20\% per time step. In short,
\begin{equation}
\Delta t^{n+1} = \min\left[ \frac{k_c}{V_{max}}, ~
                            1.2\times\Delta t^{n}
                     \right] ,
\label{eqn:timestep}
\end{equation}
where the superscript $n$ refers to the discrete time level
and the maximum velocity $V_{max}$ (computed over all zones) 
accounts for local sound speed, fluid velocity, and viscous diffusion
\begin{equation}
V_{max} = \max\left[\frac{C_s}{\min(dx^i)},~
                    \max\left(\frac{|V^i|}{dx^i}\right),~
               4k_{q2} \max(|V^i_{\ ,i}|),~
               4k_{q2} |\sum_i V^i_{\ ,i}|\right] .
\label{eqn:vmax}
\end{equation}
The Courant factor $k_{c}$ is typically set to $\lesssim 0.6$,
the viscosity strength coefficient $k_{q2}$ (defined in (\ref{eqn:av})) 
is set to $2.0$ for all the problems 
presented in this paper, and the sound speed is defined as
\begin{equation}
C_s^2 = \left.\frac{1}{h} \frac{\partial P}{\partial \rho}\right|_s
      = \frac{\Gamma P}{\rho h}
      = \frac{\Gamma(\Gamma-1) P}{(\Gamma-1)(D/W) + \Gamma P} .
\label{eqn:cs}
\end{equation}
for relativistic flows, where we have explicitly used the adiabatic eos
in the form $P\propto \rho^\Gamma$.

The artificial viscosity is implemented in a form that mimics
a general imperfect fluid stress tensor
\begin{equation}
T^{\mu\nu} = \rho h u^\mu u^\nu + P g^{\mu\nu}
           -2\eta\sigma^{\mu\nu} - \xi\theta(g^{\mu\nu} + u^\mu u^\nu) ,
\end{equation}
where $\eta$ and $\xi$ are the shear and bulk viscosity coefficients,
$\theta = u^\mu_{;\mu}$ is the expansion of fluid world lines,
and $\sigma^{\mu\nu}$ is the trace-free spatial shear tensor.
Artificial viscosity is included as a bulk scalar viscosity so
the effective perfect fluid stress energy tensor takes the form
\begin{equation}
T^{\mu\nu} = (\rho + e + P + Q) u^\mu u^\nu + (P+Q) g^{\mu\nu} ,
\end{equation}
which is equivalent to setting $P\rightarrow P+Q$ in the
momentum and energy equations
\begin{eqnarray}
 \frac{\partial S_j}{\partial t} + \frac{\partial (S_j V^i)}{\partial x^i}
 - \frac{S^\mu S^\nu}{2S^0} \frac{\partial g_{\mu\nu}}{\partial x^j}
 + \sqrt{-g} \frac{\partial (P+Q_j)}{\partial x^j} &=& 0 ,
      \label{eqn:av_momq} \\
 \frac{\partial E}{\partial t} + \frac{\partial (EV^i)}{\partial x^i}
 + \left(P+\frac{\sum_{i} Q_{i}^2}{\sum_i Q_{i}}\right)
        \frac{\partial W}{\partial t} 
 + P \frac{\partial (WV^i)}{\partial x^i} 
 + \sum_i Q_i \frac{\partial (WV^i)}{\partial x^i} 
        &=& 0 .
        \label{eqn:av_enq}
\end{eqnarray}
The scalar viscosity $Q_i$ is computed as a local quantity
in a dimensionally split fashion, and active only in convergent flows
for which $\nabla_i V^i < 0$
\begin{equation}
Q_i = (D + E + PW) \Delta l (\nabla_i V^i)
      \big[k_{q2} \Delta l (\nabla_i V^i) (1-\phi^2) - 
                  k_{q1} C_s \big].
\label{eqn:av}
\end{equation}
The coefficients $k_{q2}$ and $k_{q1}$ control the amount of
quadratic and linear (in velocity) components of viscosity, 
$\Delta l$ is a length scale set to the zone dimension,
and $\phi$ represents a limiter bounded by zero and unity that
can be applied to reduce the effect of
artificial heating and narrow the width of shock fronts.
One could alternatively use $W^{-1}\nabla_i (W V^i)$
in place of $\nabla_i V^i$ in (\ref{eqn:av}), which we find to
be effective at preventing excessively large jump errors
and helps stabilize solutions in highly relativistic
shock tube and wall shock calculations.

A more general tensor version of artificial viscosity is also
implemented for convergent flows to the form \citep{TW79}
\begin{equation}
Q_{ij} = (D + E + PW) 
         \Delta l\left[ k_{q2} \nabla_k V^k \Delta l - k_{q1} C_s\right]
         \left[\frac12(\nabla_i V^j + \nabla_j V^i) 
                            -\frac{c}{3} \nabla_k V^k \delta_{ij}\right] ,
\label{eqn:av_tensor}
\end{equation}
where $c$ is a constant defined as zero or unity depending on
whether the viscosity tensor should be traceless or not, and
$\delta_{ij}$ is the Kronecker delta.
The equations for energy and momentum with a tensor viscosity are
similar to (\ref{eqn:av_momq}) and (\ref{eqn:av_enq}) except in the
way two of the viscosity terms are computed
\begin{eqnarray}
 \frac{\partial S_j}{\partial t} + \frac{\partial (S_j V^i)}{\partial x^i}
 - \frac{S^\mu S^\nu}{2S^0} \frac{\partial g_{\mu\nu}}{\partial x^j}
 + \sqrt{-g} \frac{\partial P}{\partial x^j}
 + \sqrt{-g} \frac{\partial Q_{jk}}{\partial x^k} 
      &=& 0 , 
      \label{eqn:av2_momq} \\
 \frac{\partial E}{\partial t} + \frac{\partial (EV^i)}{\partial x^i}
 + \left(P+\frac{\sum_{i} Q_{ii}^2}{\sum_i Q_{ii}}\right)
        \frac{\partial W}{\partial t}
 +  P \frac{\partial (WV^i)}{\partial x^i} 
 +  \frac12\sum_{i,j} Q_{ij}
      \left(\frac{\partial (WV^i)}{\partial x^j} +
            \frac{\partial (WV^j)}{\partial x^i} \right)
      &=& 0 .
        \label{eqn:av2_enq} 
\end{eqnarray}
The scalar form of artificial viscosity (\ref{eqn:av}) is used in
all the tests presented in this paper.

The transport step is solved in a directionally
split, flux conservative manner. 
For example, considering advection of 
the density field along the $x$-axis in a staggered mesh scheme,
the solution to $\dot D = -\nabla_x(DV^x)$ is written
\begin{equation}
D^{n+1}_i = D^n_i - \frac{\Delta t^n}{\Delta x}
                \left[\widetilde{D}_{i+1} V_{i+1} - \widetilde{D}_i V_i\right] ,
\end{equation}
where $V_{i+1}$ is the face-centered velocity between zones $i$ and $i+1$,
and $\widetilde D_i$ is
a first order monotonic Taylor's approximation of $D_i$ from
the upwind cell center to the advection control volume center
\begin{eqnarray}
\widetilde D_i &=& \left[\frac12 + \mbox{sign}\left(\frac12,~V_i\right)\right]
  \left[D_{i-1} + \frac{(\Delta x - V_i\Delta t)}{2} (\nabla_x D)_{i-1}\right] 
  \nonumber \\
  &&\quad 
  + \left[\frac12 - \mbox{sign}\left(\frac12,~V_i\right)\right]
  \left[D_{i} - \frac{(\Delta x + V_i\Delta t)}{2} (\nabla_x D)_{i}\right] .
\label{eqn:av_upwind}
\end{eqnarray}
Equation (\ref{eqn:av_upwind}) automatically detects 
the upwind cell from the sign of the velocity $V$.
Here, $\mbox{sign}(1/2,~V_i)$ is Fortran notation for $\pm 1/2$,
depending on the sign of $V_i$.
High order \citet{VL77} monotonic interpolation
is used to reconstruct local gradients $(\nabla_x D)_i$ and prevent
spurious oscillations near regions of sharp gradients
\begin{equation}
(\nabla_x D)_i = \left[\frac12 + 
             \mbox{sign}\left(\frac12,~\Delta D_i~\Delta D_{i-1}\right)\right]
             \left(\frac{2 \Delta D_i~\Delta D_{i-1}}
                        {\Delta D_i + \Delta D_{i-1} + \delta}\right)  .
\end{equation}
The constant $\delta \ll 1$ is introduced
to prevent numerical overflow, 
and $\Delta D_i = (D_{i+1} - D_i)/\Delta x$ are the mesh aligned
gradients centered on the cell faces.
Similar expressions can easily be derived for zone-centered
variables on nonstaggered meshes by face-averaging the velocities, 
and for face-centered variables on staggered meshes by shifting the
spatial indices and control volumes appropriately.

\subsection{Non-oscillatory Central Difference Schemes}
\label{sec:central}

Considering the simplicity of equations 
(\ref{eqn:hr_de}) - (\ref{eqn:hr_mom}),
an obvious benefit of the NOCD approach is that, unlike the
AV approach, it is not expected to be
particularly sensitive to any ordering of operator updates
since the method basically just solves a 
single first order operator equation with external sources.
We have implemented two variations of this method:
the first with non-staggered spatial and temporal meshes
with second order reconstruction, and the second with
time-staggered meshes in which the variables are updated on a mesh shifted
in time to center the solution properly to second order.
A summary of the solver sequence for this class of methods is:
\begin{itemize}
\item[$\bullet$]
  Compute time step $\Delta t$ from (\ref{eqn:timestep}), \\
  redefine $\Delta t \rightarrow \Delta t/2$ for the 2-step, subcycled,
  staggered mesh scheme
\item[$\bullet$]
  Curvature
     \begin{itemize}
     \item[\bf --]
	compute pressure from the ideal fluid equation of state: \\
   	$P = (\Gamma-1)[{\cal E}\sqrt{-g}/(W^2) - D/W]
        /[\Gamma + (\Gamma-1) g^{00}(\sqrt{-g}/W)^2]$
     \item[\bf --]
        update energy and momentum, accounting for curvature: \\
	$\dot{\cal E} = \Sigma^0$ and $\dot{\cal S}^j = \Sigma^j$,
	using second order finite differencing for 
	metric derivatives
     \end{itemize}
\item[$\bullet$]
  Flux operator
     \begin{itemize}
     \item[\bf --]
	compute pressure from eos
     \item[\bf --]
        normalize velocity and update boost factor: \\
	$V^i=S^i/S^0$, using 
	$S_\mu S^\mu = -M^2 = -[({\cal E}-\sqrt{-g}g^{00}P)\sqrt{-g}/W]^2$ and \\
	$S^i = {\cal S}^i - \sqrt{-g} g^{0i} P$
	to first compute $S^0$, then the boost factor
	$W = \sqrt{-g} S^0/M$
     \item[\bf --]
	compute pressure
     \item[\bf --]
        update all variables $\omega \equiv (D,~E,~{\cal S}^j)$, \\
	accounting for flux-conservative gradient
	terms in equations (\ref{eqn:hr_de}) - (\ref{eqn:hr_mom}): \\
	$\dot \omega = -\nabla_i [\omega V^i 
	+ \sqrt{-g} P (g^{i\alpha} - g^{0\alpha} V^i)]$
     \item[\bf --]
	if the mesh is nonstaggered in time: \\
	perform interpolations to recenter variables on
	the original nonstaggered mesh \\
	$\omega^{n+1}_j = (\omega_{j-1/2}^{n+1} +
	                    \omega_{j+1/2}^{n+1})/2
	                 + (\omega_{j-1/2}^{n+1'} -
	                    \omega_{j+1/2}^{n+1'})/8$
     \end{itemize}
\item[$\bullet$]
  If the mesh is staggered: 
     \begin{itemize}
     \item[\bf --]
  	repeat curvature and flux steps to evolve solution 
	from $t=t^{n+1/2}$ to $t^{n+1}$
     \item[\bf --]
	shift array indices to realign final coordinates
	at $t^{n+1}$ with \\ 
	initial coordinates 
	at $t^n$ by $\omega_{i,j,k} = \omega_{i-1,j-1,k-1}$
     \end{itemize}
\item[$\bullet$]
  Update spacetime metric components $g_{\mu\nu}$ and
  $g^{\mu\nu}$ if time dependent
\end{itemize}

Two essential assumptions built into this method are that 
the cell-averaged solutions can be reconstructed
as MUSCL-type piece-wise linear interpolants, and that the
flux integrals are defined and evaluated naturally on staggered meshes.
Since we adopt directional splitting for multi-dimensional
problems, the basic discretization scheme used to solve
equations (\ref{eqn:hr_de}) - (\ref{eqn:hr_mom}) can be derived from a 
simple one-dimensional, first order model equation of the form
\begin{equation}
\frac{\partial\omega}{\partial t} + \frac{\partial f(\omega)}{\partial x} = 0 ,
\label{eqn:omega}
\end{equation}
where $\omega$ represents any of the density, energy or momentum variables,
and $f(\omega)$ is the associated flux.
A formal solution to (\ref{eqn:omega}) can be written over a single
time cycle $(t^n \rightarrow t^{n+1})$ on a staggered mesh as
\begin{equation}
\omega_{j+1/2}(t^{n+1}) = \omega_{j+1/2}(t^n)
-\frac{\Delta t}{\Delta x}\left[ 
    \frac{1}{\Delta t}\int_{t^n}^{t^{n+1}} f(\omega_{j+1}(\tau)) d\tau
  - \frac{1}{\Delta t}\int_{t^n}^{t^{n+1}} f(\omega_{j}(\tau)) d\tau
  \right] .
\label{eqn:formal}
\end{equation}
Introducing the notation $\omega_j' = \omega_{j+1} - \omega_{j-1}$,
the average of the piece-wise linearly reconstructed solutions
at the staggered positions $\omega_{j+1/2}(t^n)$
in (\ref{eqn:formal}) is given by
\begin{equation}
\omega_{j+1/2} = \frac12(\omega_{j+1/2}^+ + \omega_{j+1/2}^-)
  = \frac12(\omega_{j} + \omega_{j+1}) 
  + \frac{1}{8}(\omega_j'-\omega_{j+1}') ,
\label{eqn:stagger}
\end{equation}
where $\omega_{j+1/2}^\pm$ refer to the piecewise linearly interpolated
solutions from the upwind and downwind cell centers
\begin{eqnarray}
\omega_{j+1/2}^+ &=& \omega_{j+1} - \frac{1}{4}(\omega_{j+2}-\omega_j), \\
\omega_{j+1/2}^- &=& \omega_{j} + \frac{1}{4}(\omega_{j+1}-\omega_{j-1}).
\end{eqnarray}

Considering that the time averaged
integrals in (\ref{eqn:formal}) can be approximated using midpoint values
\begin{equation}
\frac{1}{\Delta t}\int_{t^n}^{t^{n+1}} f(\omega_{j}(\tau)) d\tau
  \sim f(\omega_j(t^{n+1/2})) ,
\end{equation}
immediately suggests a two step predictor-corrector procedure
to solve (\ref{eqn:omega}): the state variables are predicted
at $t=t^{n+1/2}$ by
\begin{equation}
\omega_j^{n+1/2} = \omega_j^n 
                 - \frac{\Delta t}{2\Delta x} f'(\omega_j) ,
\label{eqn:pred}
\end{equation}
then corrected on the staggered mesh by
\begin{equation}
\omega_{j+1/2}^{n+1} = 
  \frac12(\omega_j^n + \omega_{j+1}^{n})
 +\frac18(\omega_j' - \omega_{j+1}')
 -\frac{\Delta t}{\Delta x} \left[ f(\omega_{j+1}^{n+1/2})
                                  -f(\omega_{j}^{n+1/2})
                            \right] ,
\label{eqn:corr}
\end{equation}
where we have also substituted (\ref{eqn:stagger}) for
$\omega_{j+1/2}(t^n)$ in (\ref{eqn:formal}).
Equations (\ref{eqn:pred}) and (\ref{eqn:corr}) represent the
complete single cycle solution averaged on a staggered mesh. The
mesh indices can be brought back into alignment by setting
$\Delta t \rightarrow \Delta t/2$, performing two time cycle
updates (computing $\omega_{j+1/2}^{n+1/2}$ then
$\omega_{j+1}^{n+1}$)
to time $t^{n+1} = t^n + \Delta t$, and re-center
the solution on the original zone positions by shifting
the array indices as $\omega_j = \omega_{j-1}$.

As an alternative to mesh staggering, the solution after
applying the corrector step can be reconstructed directly
back to the nonstaggered cell centers by a second order
piece-wise linear extrapolation
\begin{equation}
\omega_{j}^{n+1} 
  = \frac12(\omega_{j-1/2}^{n+1} + \omega_{j+1/2}^{n+1}) 
  + \frac{1}{8}(\omega_{j-1/2}^{n+1'}-\omega_{j+1/2}^{n+1'}) ,
\end{equation}
to yield for the final single time-step solution
\begin{eqnarray}
\omega_j^{n+1} &=& \frac14(\omega_{j-1}^{n} + 
      2\omega_{j}^{n} + \omega_{j+1}^{n})
    - \frac{1}{16}\left( \omega_{j+1}^{n'} - \omega_{j-1}^{n'}\right) 
    \nonumber \\
    &&\quad - \frac{\Delta t}{2\Delta x}\left[f(\omega_{j+1}^{n+1/2}) - 
                                      f(\omega_{j-1}^{n+1/2})\right]
    - \frac{1}{8}\left( \omega_{j+1/2}^{n+1'} - \omega_{j-1/2}^{n+1'}\right) .
\label{eqn:nonstag}
\end{eqnarray}
We have found no substantial differences between the staggered
and nonstaggered approaches in all the test calculations we have
performed. Hence all subsequent results
presented in this paper from this class of algorithms are
derived with the nonstaggered mesh method using
(\ref{eqn:pred}) and (\ref{eqn:nonstag}).

One final important element of this method is that all gradients
(of either the state variables $\omega_j'$ or fluxes
$f'(\omega_j)$) must be processed and limited for monotonicity
in order to guarantee non-oscillatory behavior.
This is accomplished with either the minmod limiter
\begin{equation}
\omega_j'= \max\left(0,~\min\left(1,~
           \frac{\omega_j - \omega_{j-1}}{\omega_{j+1} - \omega_j}
           \right)\right) \left(\omega_{j+1} - \omega_j\right) ,
\end{equation}
or the van Leer limiter
\begin{equation}
\omega_j'= \left[\frac{  |(\omega_j - \omega_{j-1})/(\omega_{j+1}-\omega_j)| +
                         (\omega_j - \omega_{j-1})/(\omega_{j+1}-\omega_j)}
                      {1+|(\omega_j - \omega_{j-1})/(\omega_{j+1}-\omega_j)|}\right]
           \left(\omega_{j+1} - \omega_j\right) ,
\end{equation}
which satisfy the TVD constraints with appropriate 
Courant restrictions, although we note that steeper limiters
can yield undesirable results especially in under-resolved
high boost shock tube calculations.

\section{Code Tests}
\label{sec:tests}

\subsection{Relativistic Shock Tube}
\label{sec:stube}

We begin testing the staggered AV and 
nonstaggered NOCD methods with one of the standard problems
in fluid dynamics, the shock tube.  This test is characterized 
initially by two different fluid states separated by a membrane.
At $t=0$ the membrane is removed and the fluid evolves in 
such a way that five distinct regions appear in the flow: an 
undisturbed region at each end, separated by a rarefaction wave, 
a contact discontinuity, and a shock wave.  This problem only 
checks the hydrodynamical elements of the code, as it assumes 
a flat background metric.  However, it provides a good test 
of the shock-capturing properties of the different methods since
it has an exact solution \citep{Thompson86}
against which the numerical results can be compared.

Two cases of the shock tube problem are considered first:
moderate boost ($W=1.43$) and high boost ($W=3.59$) shock waves.
In the moderate boost case, the 
initial state is specified by $\rho_L = 10$, $P_L = 13.3$, and 
$V_L = 0$ to the left of the membrane and $\rho_R = 1$, 
$P_R = 10^{-6}$, and $V_R = 0$ to the right.  In the high boost 
case, $\rho_L = 1$, $P_L = 10^3$, $V_L = 0$, and $\rho_R = 1$, 
$P_R = 10^{-2}$, $V_R = 0$.  In both cases, the fluid is 
assumed to be an ideal gas with $\Gamma = 5/3$, and the 
integration domain extends over a unit grid
from $x=0$ to $x=1$, with the membrane located at $x=0.5$.
The AV shock tube results presented here were run
using the scalar artificial viscosity with a quadratic
viscosity coefficient $k_{q2}=2.0$, linear viscosity coefficient
$k_{q1}=0.3$, and Courant factor $k_c=0.6$ (0.3 for the highest boost cases).
For the NOCD method we use $k_c=0.3$ and the minmod limiter which gives
smoother and more robust results than the steeper limiters
in simulations of under-resolved highly relativistic shocks.
We have carried
out these tests in one, two and three dimensions, lining
up the interface membrane along the main diagonals in
multi-dimensional runs.
We found it necessary, in order to maintain stability 
in the highest boost cases,
to impose constraints on the pressure and energy density 
at each cycle to ensure they maintained positive values.
Although this wasn't necessary at velocities smaller than about 0.95, we
nevertheless enforced these conditions in all the calculations.

\begin{figure}[htb]
\plotone{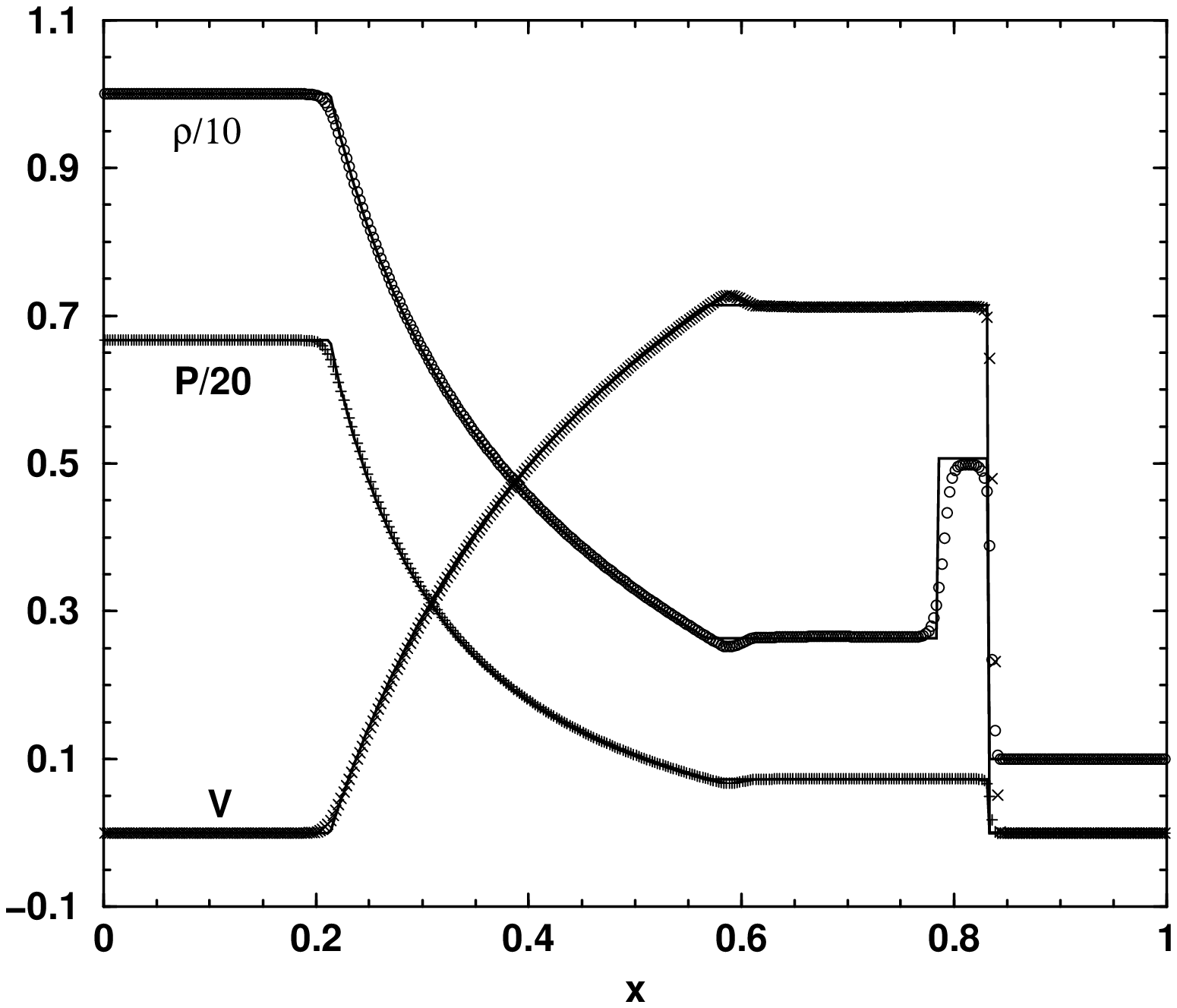}
\caption{Normalized results for the moderate boost shock tube test using 
artificial viscosity for shock capturing and 400 zones 
to cover a unit grid.  The solution is shown at time $t=0.4$.}
\label{fig:fig1}
\end{figure}

\begin{figure}[htb]
\plotone{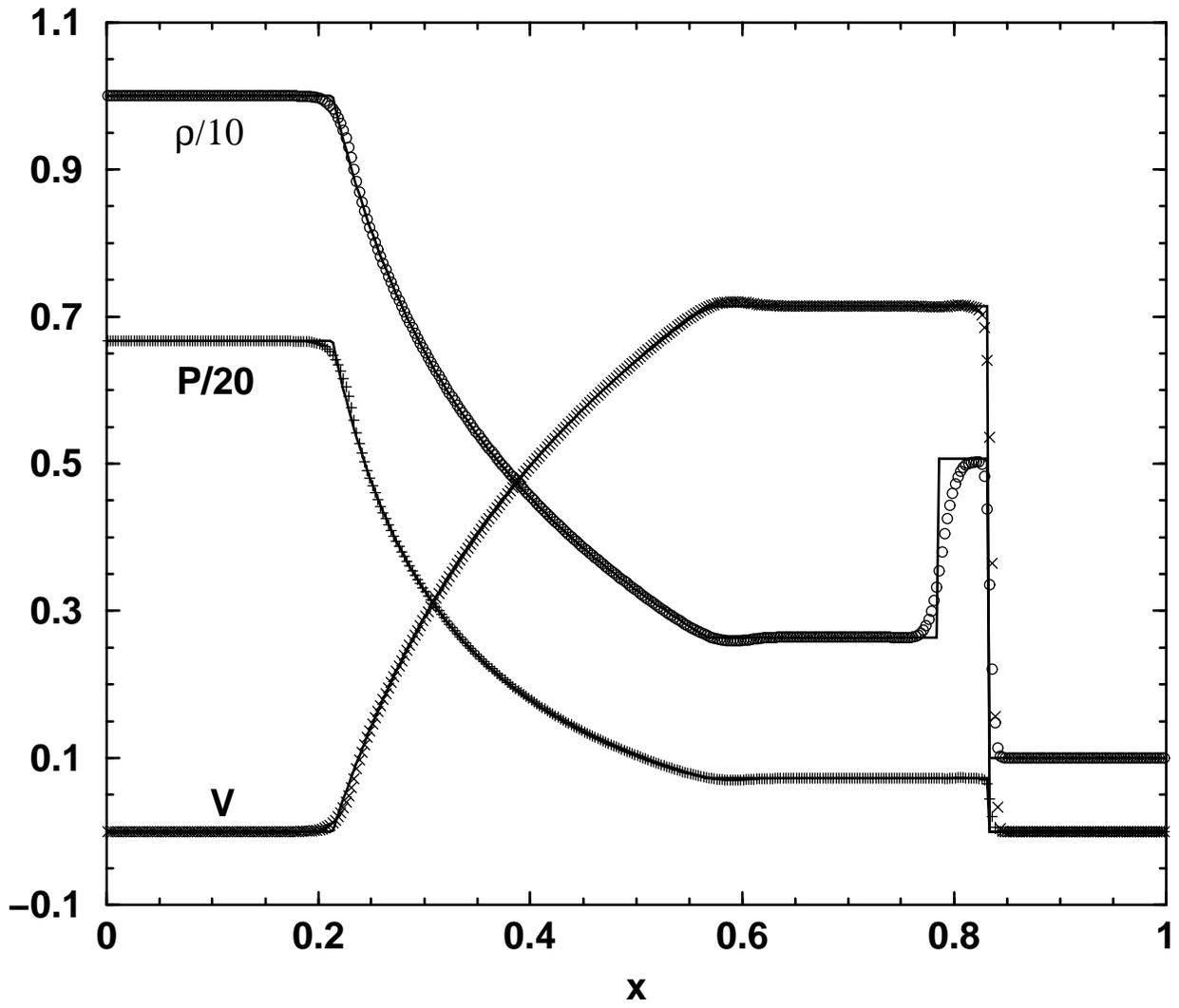}
\caption{As Figure \protect{\ref{fig:fig1}} except the solutions are computed
using the non-oscillatory central difference scheme.}
\label{fig:fig2}
\end{figure}

\begin{figure}[htb]
\plotone{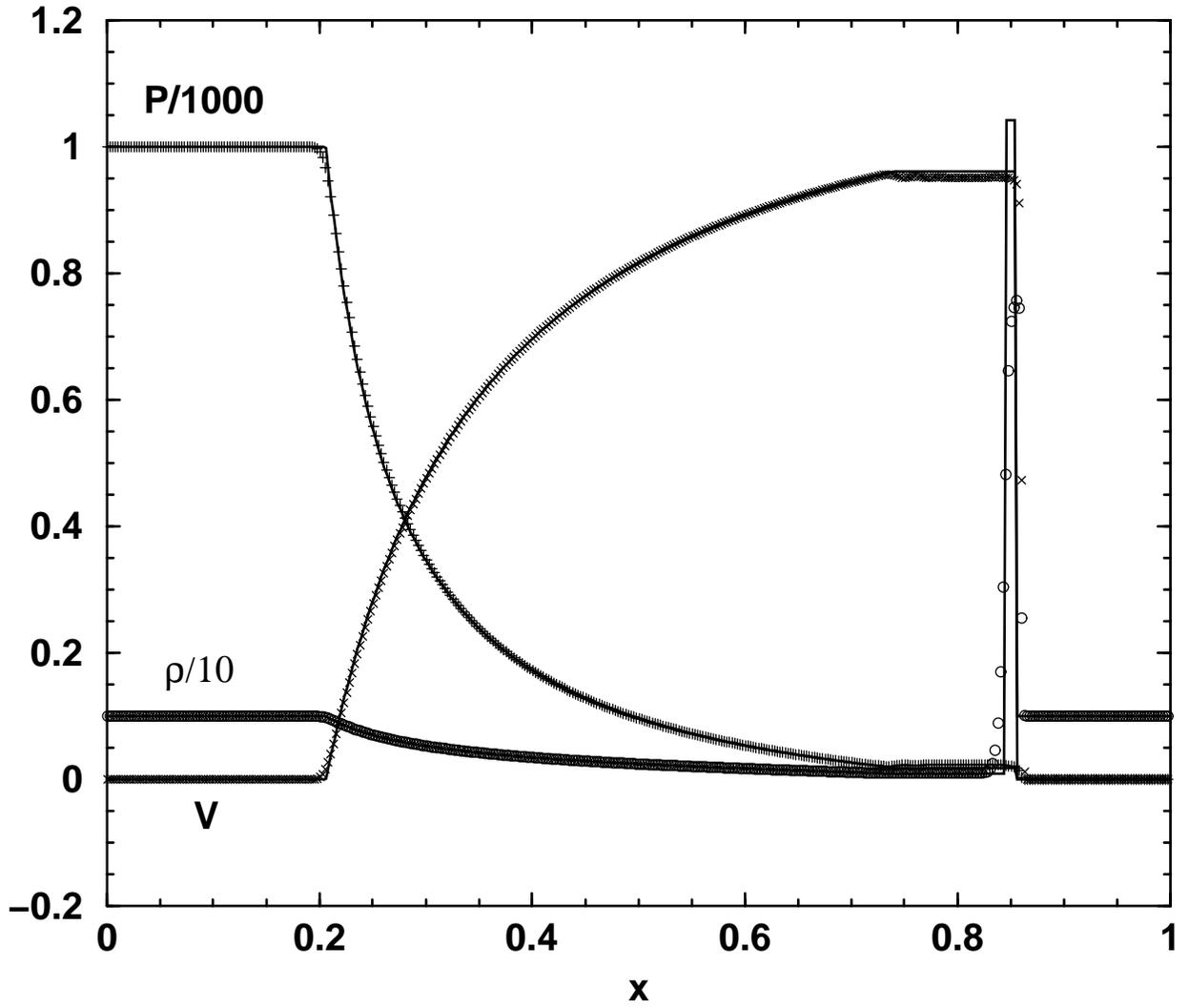}
\caption{Results at time $t=0.36$
for the high boost shock tube test using artificial
viscosity and 800 zones.  The data points in this plot have been
sampled to reduce overcrowding.  Only 400 points are shown.}
\label{fig:fig3}
\end{figure}

\begin{figure}[htb]
\plotone{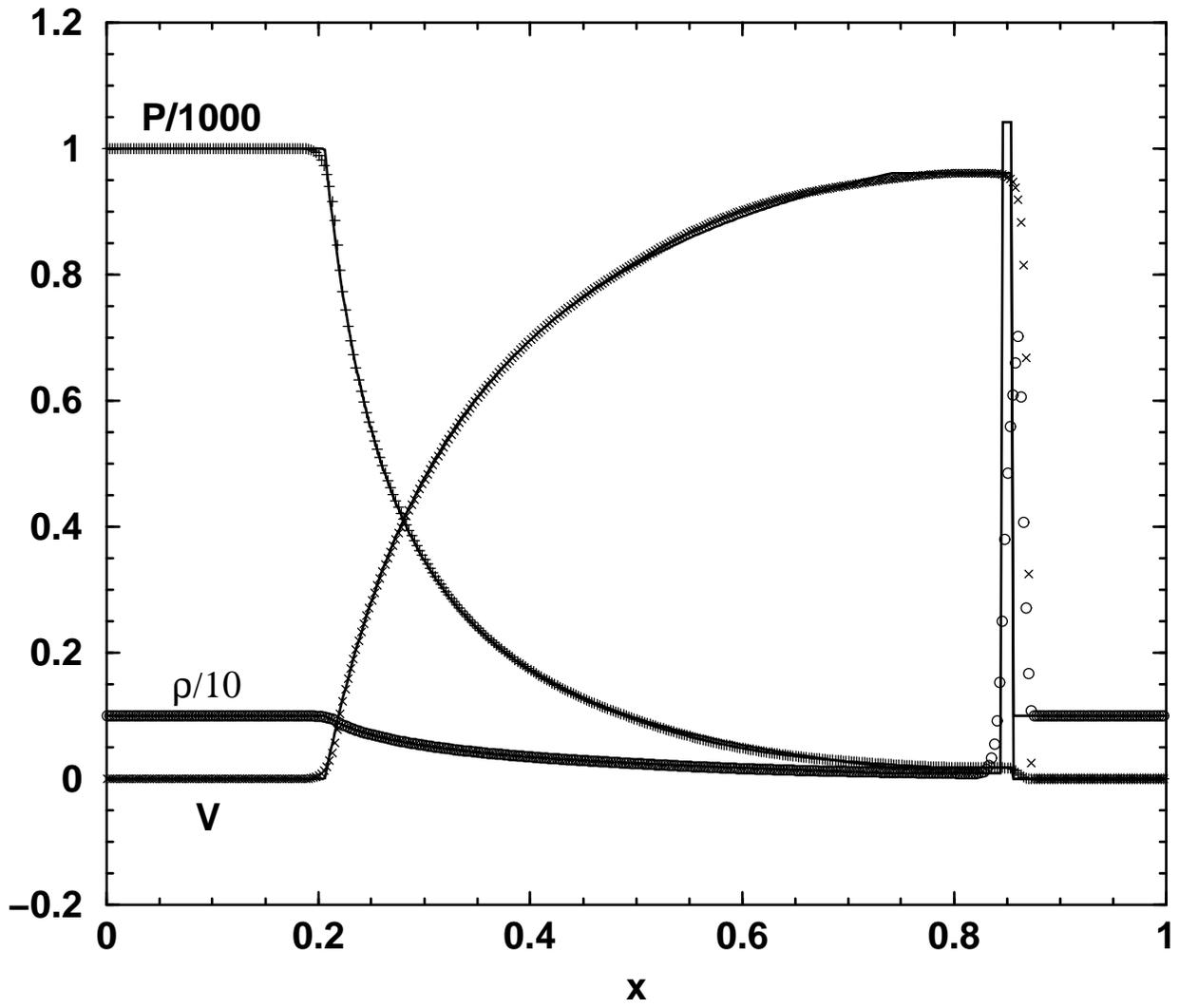}
\caption{As Figure \protect{\ref{fig:fig3}} but with
the non-oscillatory central difference scheme.}
\label{fig:fig4}
\end{figure}

Figures \ref{fig:fig1} \& \ref{fig:fig2} show 
spatial profiles of the moderate boost results at time $t=0.4$
on a grid of 400 uniformly spaced zones using the AV and NOCD 
methods respectively.
Figures \ref{fig:fig3} \& \ref{fig:fig4}
show the corresponding solutions of both AV and NOCD methods for
the high boost test
using a higher resolution grid with 800 zones at time $t=0.36$.
The density under-shoot (about 30\%) in Figures \ref{fig:fig3}
and \ref{fig:fig4} is due to a lack of
sufficient spatial resolution, and improves significantly
by increasing the number of zones.
Tables \ref{tab:errors1} \& \ref{tab:errors2} summarize the errors in the
primitive variables $\rho$, $P$, and $V$ for different grid resolutions
and CFD methods using the $L$-1 norm (i.e.,
$\Vert E(a) \Vert_1 = \sum_{i,j,k} \Delta x_i \Delta y_j \Delta z_k
\vert a_{i,j,k}^n - A_{i,j,k}^n \vert$, where
$a_{i,j,k}^n$ and $A_{i,j,k}^n$ are the numerical and exact 
solutions, respectively, and for 1D problems the orthogonal grid spacings
are set to unity).
Included in Table \ref{tab:errors1} for comparison are the
errors reported by \citet{Font00} using Marquina's
approximate Riemann solver \citep{Donat96}.  They also tested
the Roe and Flux-split approximate solvers and achieved similar results
as with Marquina's method, so we do not include those numbers here.
We find the errors in Table \ref{tab:errors1} are quite comparable 
between all three methods
with convergence rates just under first order as expected
for shock capturing methods.
For the high boost case in Table \ref{tab:errors2}, 
our errors are comparable to those reported by
\citet{Marti96} for
the same shock tube simulation using
an extended high order piecewise parabolic method (PPM)\citep{Colella84}
with an exact Riemann solver.
However, we note that their published errors are for
the conserved quantities (generalized fluid density $D$, 
generalized energy density $\tau = \rho h W^2 - P - D$, 
and covariant momentum density $S$) rather than the primitive variables
we report.  Their results are included in Table \ref{tab:errors2}.
We also note that the slightly larger errors in the 3D AV results
of Table \ref{tab:errors1} are due
primarily to boundary effects (particularly at the grid corners) 
and not to shock capturing differences.
In fact, errors computed only along the main
diagonal are about the same for the NOCD
and AV methods.

\begin{figure}[htb]
\plotone{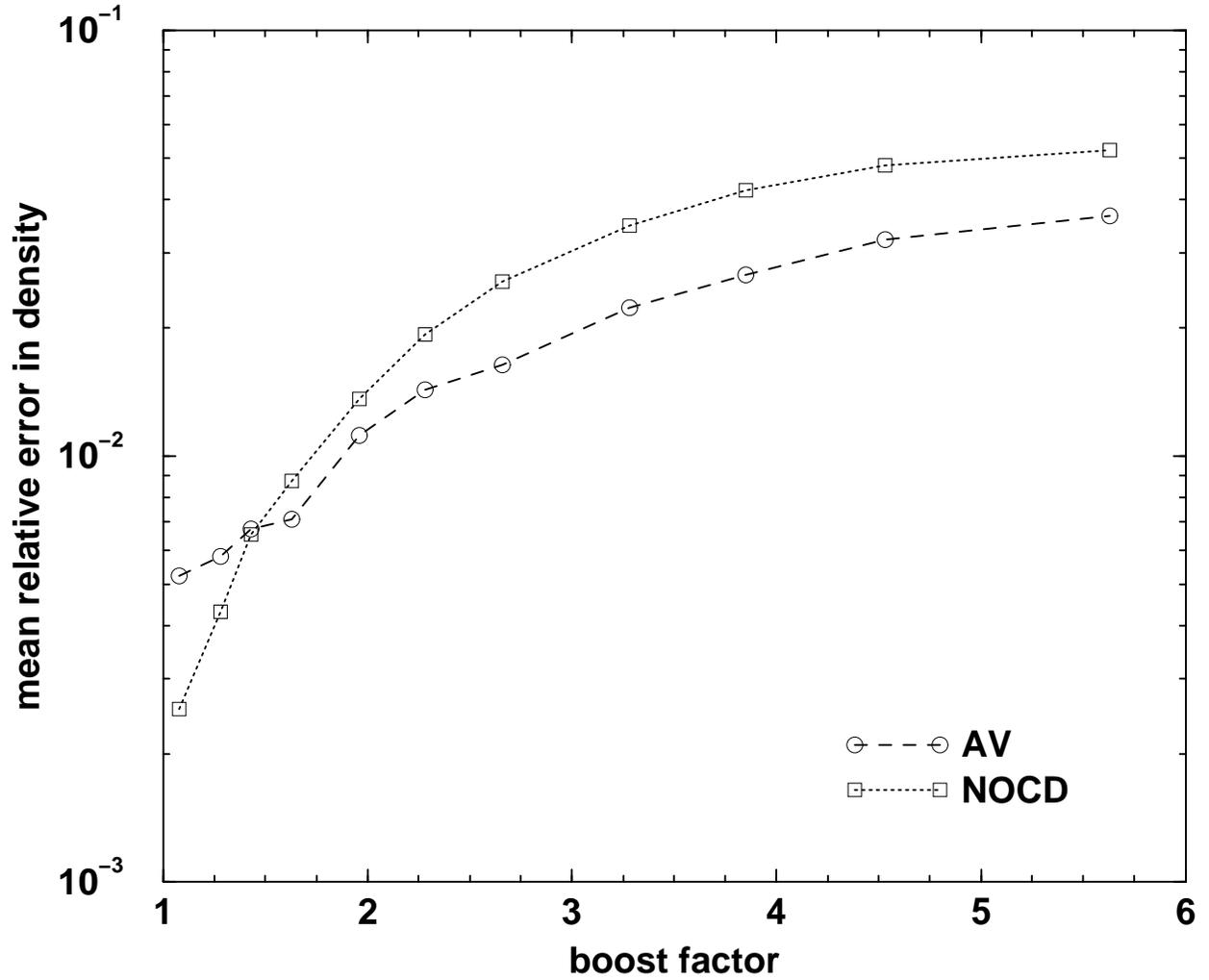}
\caption{Mean relative errors in density for both the AV
and NOCD methods as a function of boost for the relativistic shock tube 
problem. All calculations were run using 800 zones up to time $t=0.4$.
The maximum boost limit shown in this plot
(corresponding to a velocity of 0.984) 
is set by the constraint to resolve the
leading post-shock density plateau by at least four cells
in the analytic solution.
}
\label{fig:fig5}
\end{figure}

Table \ref{tab:errors3} shows the mean-relative errors
(defined as $\bar{\epsilon}_\mathrm{rel}(a)
= \sum_{i,j,k} \vert a_{i,j,k}^n - A_{i,j,k}^n \vert / \sum_{i,j,k} 
\vert A_{i,j,k}^n \vert$,
where again $a_{i,j,k}^n$ and $A_{i,j,k}^n$ are the numerical and exact 
solutions, respectively) in the primitive variables over a 
range of boost factors
using 800 zones to cover the same unit domain. The different boost
factors are established
by systematically increasing the original value of $P_L$ over
the moderate boost case. These errors are also displayed graphically
in Figure \ref{fig:fig5}, comparing the AV and NOCD methods up to
the maximum boost ($W=5.63$ corresponding to a velocity of
$V=0.984$) allowed at this grid resolution, which we define
as four cells to cover the leading post-shock density plateau
using the analytic solution as a guide.
The increasing trend (with boost) in error
reflects the stronger nonlinear coupling through the fluid velocity
and the narrower and steeper leading shock plateau found
in the density plots of Figures \ref{fig:fig3} and \ref{fig:fig4}. 
Over the range of shock velocities we have simulated,
the errors are comparable between the AV, NOCD, and Godunov methods.

\subsection{Relativistic Wall Shock}
\label{sec:wallshock}

A second test presented here is the wall shock problem involving
the shock heating of cold fluid hitting a wall at the left boundary
($x=0$) of a unit grid domain. The initial data are set up to be
uniform across the grid with adiabatic index $\Gamma=4/3$, 
pre-shocked density $\rho_{1} = 1$,
pre-shocked pressure $P_{1} = 7.63 \times 10^{-6}$, and 
velocity $V_{1} = -v_{init} = -(1-\nu)$, where $\nu < 1$
is the infall velocity parameter. When the fluid hits the wall
a shock forms and travels to the right, separating the pre-shocked
state composed of the initial data and the post-shocked state
with solution in the wall frame
\begin{eqnarray}
V_S      & = & \frac{\rho_{1} W_{1} V_{1}}{\rho_{2} - \rho_{1}W_{1}} , \\
P_{2}    & = & \rho_{2} (\Gamma - 1)(W_{1} - 1) , \\
\rho_{2} & = & \rho_{1} \left[ \frac{\Gamma + 1}{\Gamma - 1} +
               \frac{\Gamma}{\Gamma - 1}(W_{1} - 1) \right] , 
\end{eqnarray}
where $V_S$ is the velocity of the shock front, and the pre-shocked
energy and post-shocked velocity were both assumed
negligible ($\epsilon_1 = V_2 = 0$).
To facilitate a direct comparison between our results and the Genesis
code of \citet{Aloy99} (which again uses
Marquina's approximate Riemann solver) 
all of the results shown in the figures and tables, unless noted otherwise,
were performed on a 200 zone uniformly spaced mesh and ran to a final
time of $t=2.0$. Also, for the NOCD methods, the Courant factor is set to $k_c = 0.3$,
and we use the van Leer limiter for gradient calculations, which
generally gives smaller errors when compared to the
more diffusive minmod limiter (about a 30\% reduction for the lower boost cases
we have tried). For the AV methods, we use the scalar viscosity with $k_c = 0.6$,
$k_{q1} = 0.3$, and $k_{q2} = 2.0$ for all the runs.

\begin{figure}[htb]
\plotone{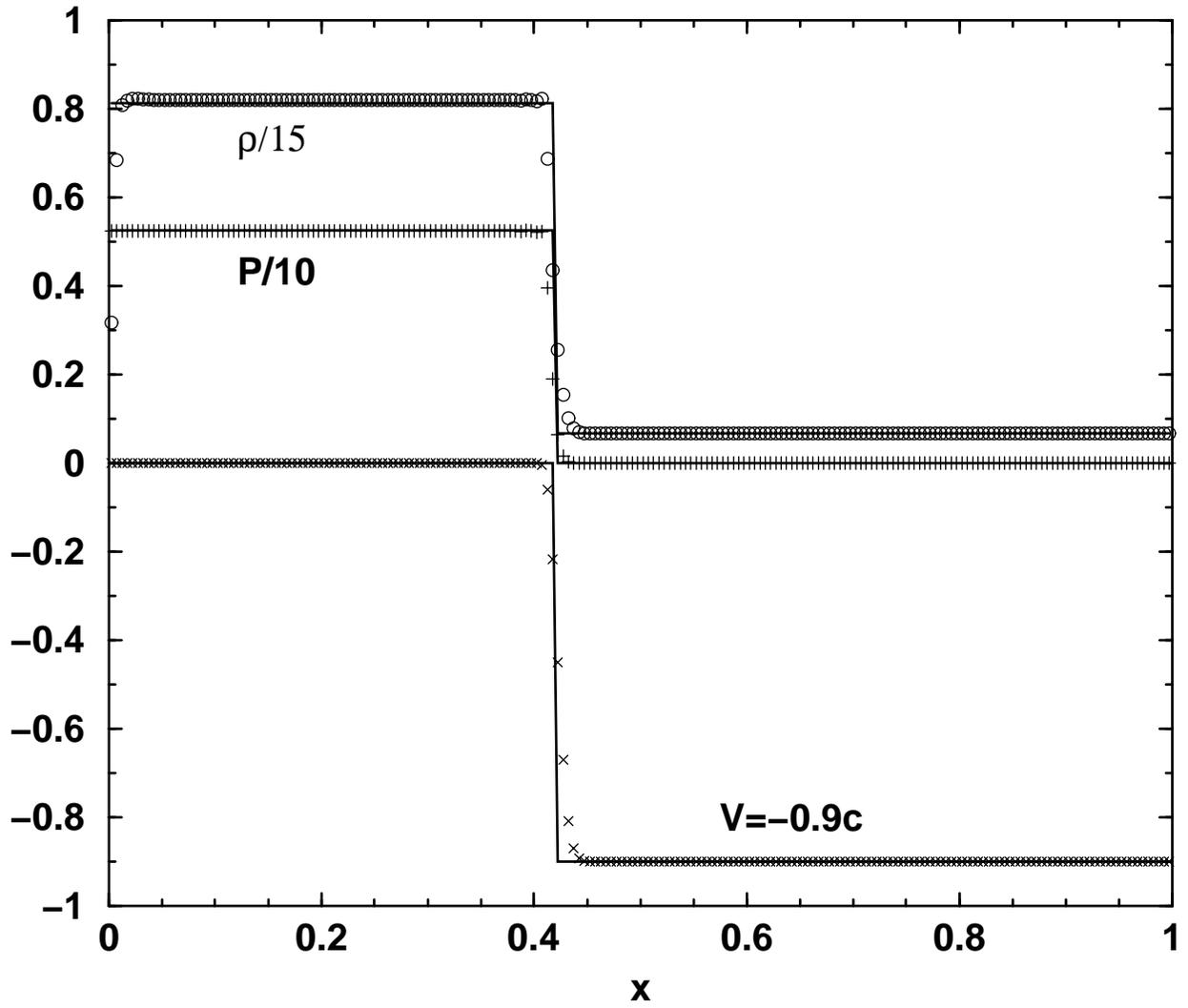}
\caption{Results for the relativistic
wall shock test with 200-zone resolution and infall
velocity $V=-0.9c$ using artificial viscosity.  The solution is shown at time $t=2.0$.}
\label{fig:fig6}
\end{figure}

\begin{figure}[htb]
\plotone{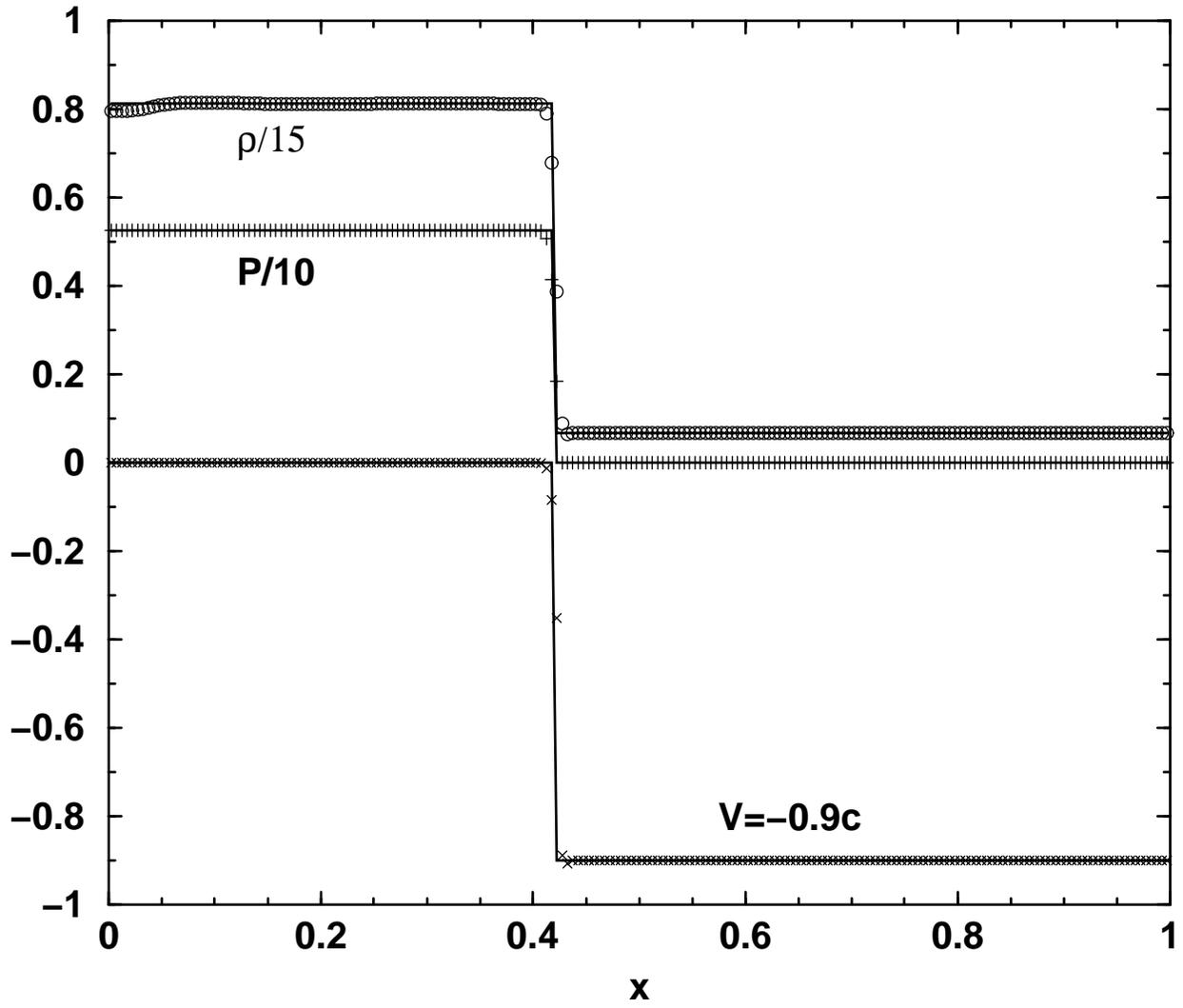}
\caption{As Figure \protect{\ref{fig:fig6}} but for the NOCD scheme.}
\label{fig:fig7}
\end{figure}

Figures \ref{fig:fig6} \& \ref{fig:fig7} show spatial profiles for
the case with initial velocity $v_{init} = 0.9$ and 200 zones for the AV and 
NOCD methods, respectively. Table \ref{tab:errors4} 
summarizes the $L$-1 norm errors in both methods
as a function of grid resolution.
The values given in parentheses 
are the contributions to the total error in the first twenty
zones from the reflection wall at $x=0$. These numbers clearly indicate
a disproportionate error distribution from wall heating, an effect that
is especially evident in the AV results, and particularly in the density
curve where the first two data points in Figure \ref{fig:fig6} differ
significantly from the true post-shock state.
Excluding this contribution may give a more accurate
assessment of each method's ability to resolve the actual shock profile. 

\begin{figure}[htb]
\plotone{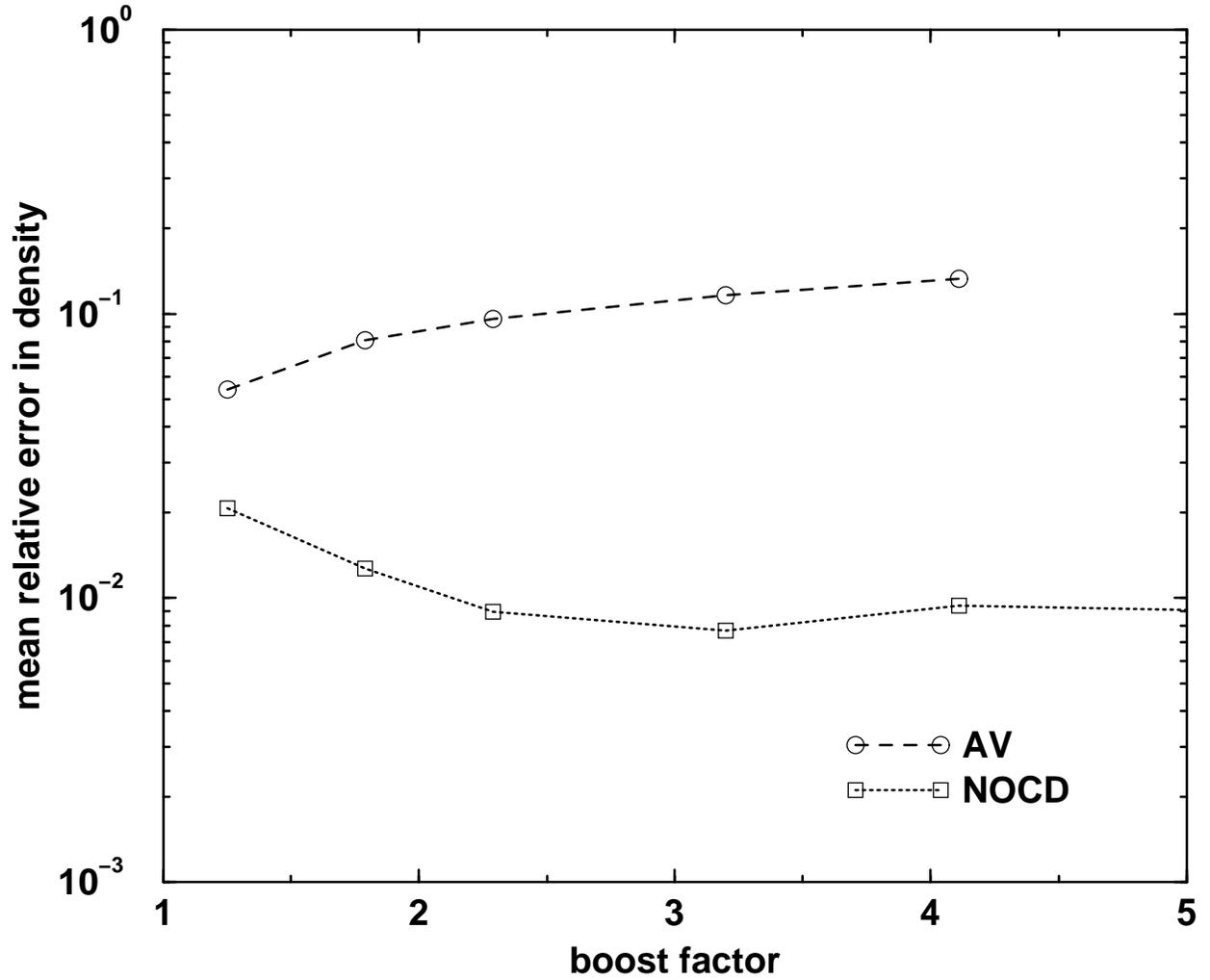}
\caption{Mean relative errors in density for both the AV
and NOCD methods as a function of boost for the relativistic wall shock 
problem. All calculations were run using 200 zones up to time $t=2.0$.
The AV results can be significantly improved
and brought closer in alignment with the NOCD results
by simply reducing the Courant factor or increasing the viscosity
coefficients over the standard values we have chosen for all the tests.
}
\label{fig:fig8}
\end{figure}

\begin{figure}[htb]
\plotone{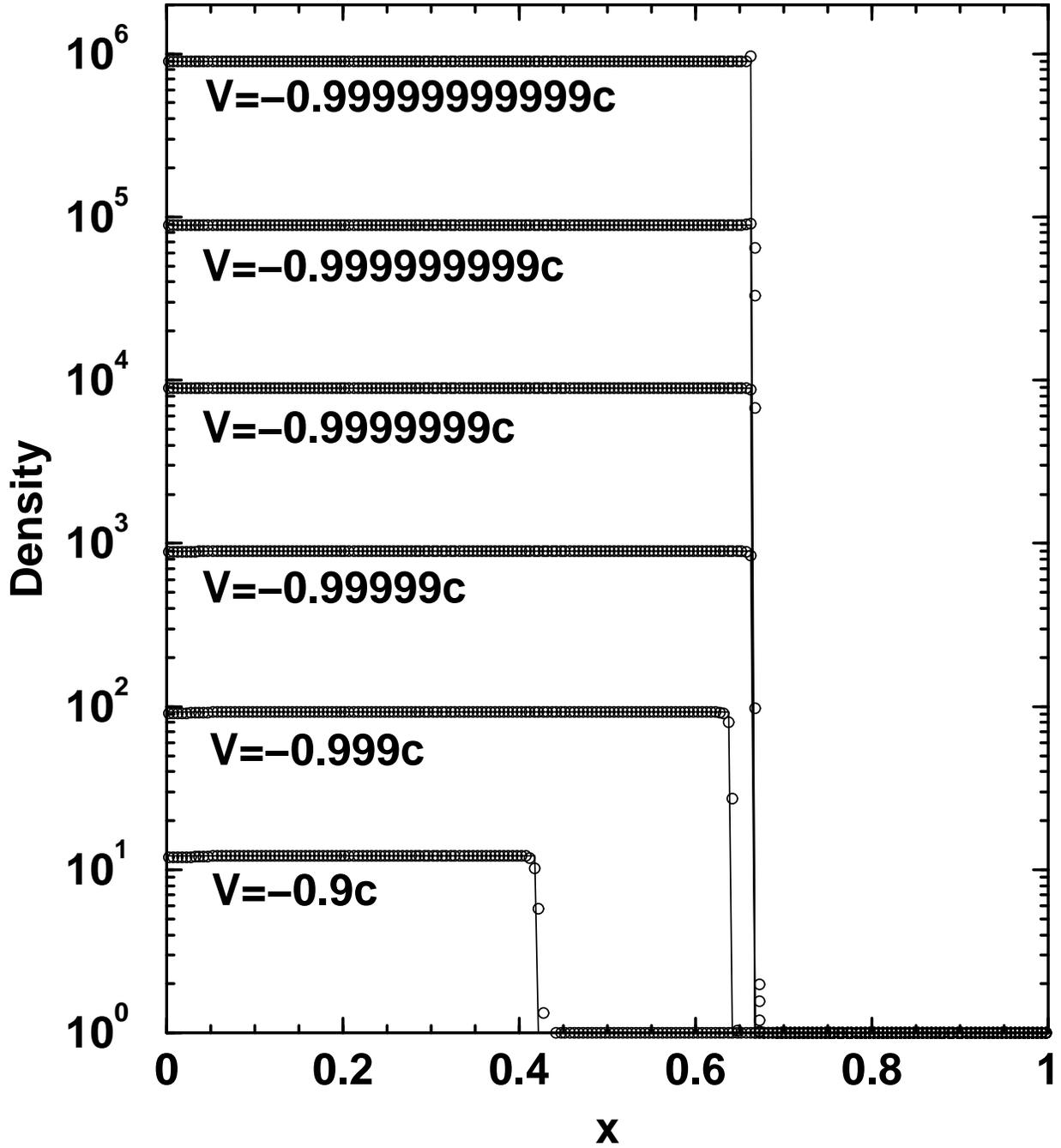}
\caption{Density plots for different infall velocities in the 
wall shock test using the NOCD method.  The resolution 
is 200 zones and the displayed time is $t=2.0$.}
\label{fig:fig9}
\end{figure}

Figure \ref{fig:fig8} plots the mean-relative errors (using 200 zones)
in density, which are generally greater than errors
in either the pressure or velocity, 
as a function of boost factor up to about the maximum boost
that the AV methods can be run accurately.
Although we are not able to extend the AV method reliably 
(which we define by a 10\% mean error threshold, and increased
sensitivity to viscosity parameters)
beyond $v_{init} \sim 0.97$, the NOCD methods, on the other hand, are
substantially more robust. In fact, as shown in Table \ref{tab:errors5}
and Figure \ref{fig:fig9}, the NOCD schemes can be run up to arbitrarily high
boost factors with stable mean relative errors, typically
less than two percent with no significant increasing trend.
These errors are generally smaller than those quoted by \citet{Aloy99}.
However, we note that the errors for the AV method presented in
Figure \ref{fig:fig8} and Table \ref{tab:errors5} can be improved significantly
by either lowering the Courant factor or increasing the viscosity
coefficients. For example, decreasing $k_c$ from
0.6 to 0.3, or increasing $k_{q2}$ from 2 to 3 for the case $v_{init}=0.95$
reduces the $L$-1 norm in density from 0.116 to 0.048 and 0.033, respectively.
We have also been able to run accurate wall shock tests with the AV method
at higher boosts than shown in Table \ref{tab:errors5} by choosing
different parameter combinations (e.g., $k_c=0.2$, $k_{q1} = 1.0$
and 400 zones can evolve flows with $v_{init} = 0.99$ fairly well).
However, rather than adjusting parameters to achieve the best possible result 
for each specific problem, we have opted to keep numerical parameters constant 
between code tests, boost factors, and numerical methods.

\subsection{Black Hole Accretion}
\label{sec:accretion}

As a test of hydrodynamic flows in spacetimes with nontrivial
curvature, we consider
radial accretion of an ideal fluid onto a compact,
strongly gravitating object, in this case a 
Schwarzschild black hole.
The fluid will accrete onto the compact object along geodesics, thus allowing
the general relativistic components of our codes to be tested against a 
well-known analytic stationary solution.
Assuming a perfect fluid in isotropic Schwarzschild coordinates
\begin{equation}
ds^2 = -\left(\frac{1-M/2\rho}{1+M/2\rho}\right)^2 dt^2
       +\left(1+\frac{M}{2\rho}\right)^4
        \left(dx^2 + dy^2 + dz^2 \right) ,
\end{equation}
where $\rho = \sqrt{x^2 + y^2 + z^2}$ is the isotropic radius,
the exact solution to this problem is dependent on a single parameter, the
gravitational binding energy ($u_0$).  In terms of this parameter
(which we set to $u_0=-1$ in our tests),
the solution can be written
\begin{eqnarray}
W      & = & \sqrt{-g}~g^{00} u_0 , \\
D      & = & \frac{C_D}{\rho^2 V^\rho} , \\
E      & = & \frac{C_E}{W^{\Gamma -1} (\rho^2 V^\rho)^\Gamma} , \\
V^\rho & = & -\frac{u^\rho}{u^0} = -\frac{1}{g^{00} u_0} 
              \sqrt{ -g^{\rho\rho}(1+g^{00} u_0^2)} ,
\end{eqnarray}
where $W$ is the boost factor,
$V^\rho$ is the radial infall velocity in isotropic
radial coordinates, $D$ is the generalized
density in isotropic Cartesian coordinates, 
$E$ is the generalized internal energy in isotropic Cartesian coordinates, 
$\Gamma=5/3$ is the adiabatic index, and
$C_D$ and $C_E$ are constants of integration which
we set to $C_D=1$ and $C_E = C_D/100$ in the simulations.

\begin{figure}[htb]
\plotone{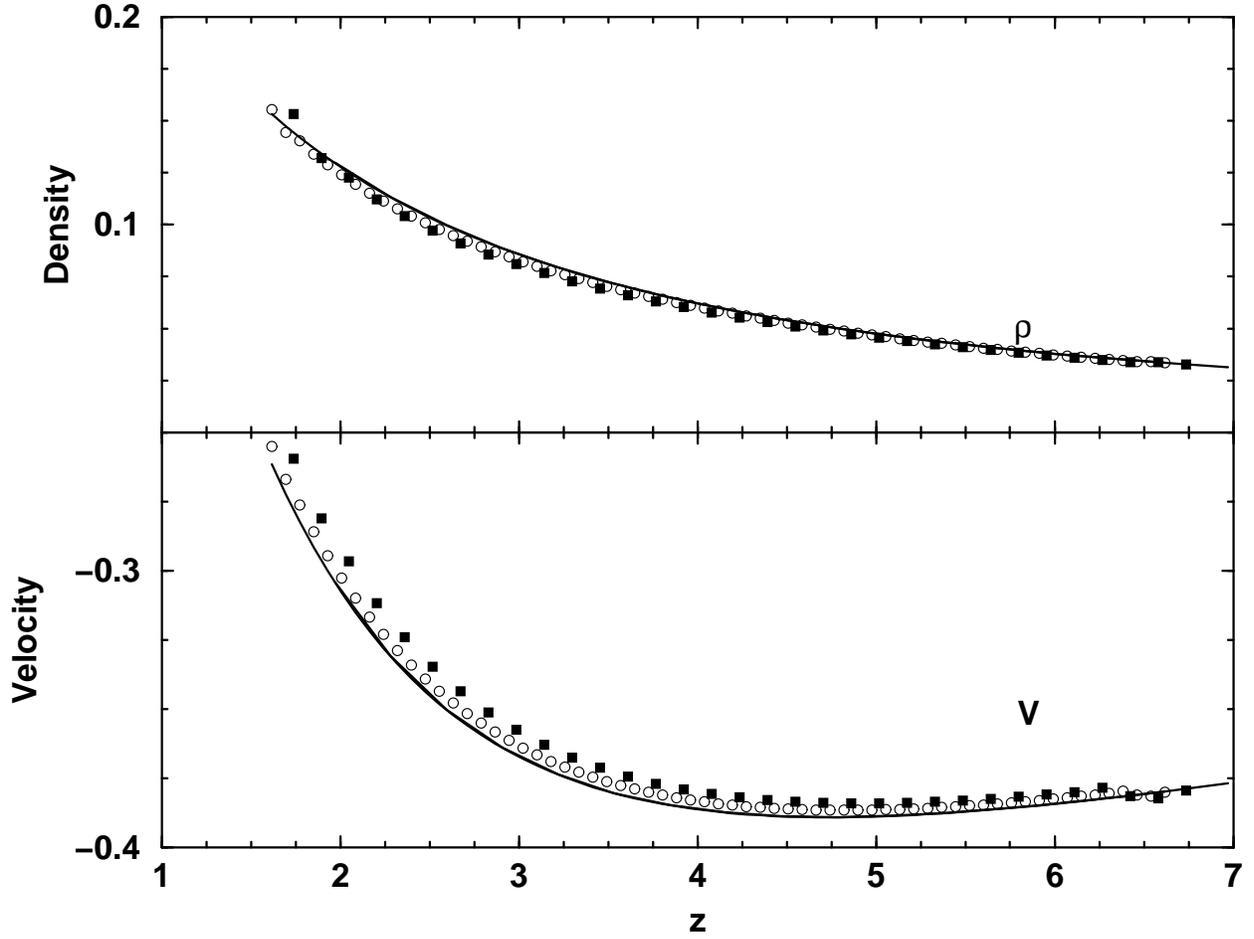}
\caption{Plots of density and velocity along the $z$-axis
for the dust accretion problem using the AV method.  The filled squares
and open circles correspond to resolutions of $32^3$ and $64^3$, respectively.
The solid line is the analytic solution.  The displayed time is $t=50M$.
}
\label{fig:fig10}
\end{figure}

\begin{figure}[htb]
\plotone{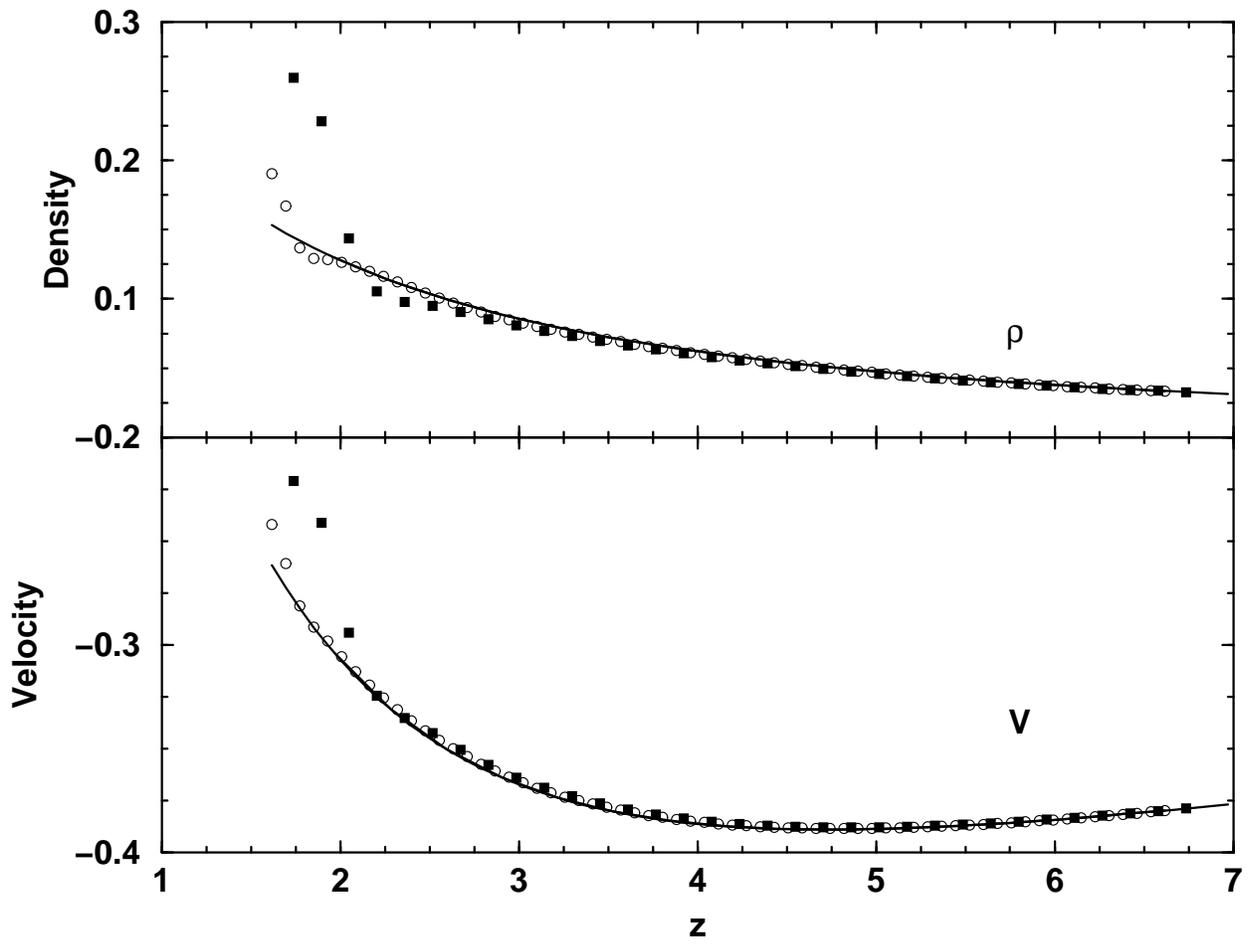}
\caption{As Figure \protect{\ref{fig:fig10}}, except for the NOCD method.
}
\label{fig:fig11}
\end{figure}

The computational domain for this problem is constructed to be $(5 M)^3$
(where $M=1$ is the black hole mass)
and centered along the $z$-axis
with $-2.5M \le (x,~y) \le 2.5M$. In the $z$-direction the
inner boundary zone $z=z_{min}$ is defined to lie outside the event
horizon at $z_{min} = 1.5M$ in isotropic coordinates to
guarantee all boundary zones are outside the horizon, and extends
to $z=z_{max}=6.5M$ along the $x=y=0$ line.
Calculations are carried out on different resolution grids,
ranging from $16^3$ to $64^3$ to check code convergence.
All variables are initially set to negligible values throughout the
interior domain ($D=10^{-2}$, $E=D C_E/C_D$, $W=1$, and $S_i=V^i=0$),
and the static analytic solutions are specified as outer boundary
conditions at all times.
Along the inner $z$ boundary,
outflow conditions are maintained by simply setting the
first derivatives of all variables to zero at the end of each time step.
Thus fluid flows onto the computational grid from all of the
analytically-specified (inflow) boundaries, and exits from
the lower $z$-plane closest to the black hole.
All results presented here were generated from simulations
run until steady-state was achieved at $t=50M$, and
numerical parameters are defined as in previous tests, namely
$k_c=0.6$, $k_{q1}= 0$, $k_{q2}=2.0$ in the AV runs,
and $k_c=0.3$ in the NOCD results. Table \ref{tab:errors6}
summarizes the global mean-relative errors in both methods
as a function of grid resolution. Figures \ref{fig:fig10}
\& \ref{fig:fig11} show spatial profiles of density and velocity along
the $z$-axis for $32^3$ and $64^3$ zones for the AV and 
NOCD methods, respectively.  

Although the numerical results in Table \ref{tab:errors6}
converge to the analytic solution with grid resolution, they converge
at a rate between first and second order due in part to the treatment of
boundary conditions and time discretization errors. In particular,
comparing the analytic and numerical
solutions, we find that maximum relative errors occur near
the event horizon along the inner $z$-boundary.  
For the AV method, the maximum relative errors for
density and velocity with $64^3$ zones
are 9.16\% and 2.49\%, respectively, compared to global
mean-relative errors of 1.36\% and 0.63\%.  For the NOCD method, the maximum
relative errors are 24.4\% and 7.42\%, compared to global mean-relative 
errors of 2.11\% and 0.14\%.  The global errors in both methods,
in spite of being computed on a nonsymmetric Cartesian mesh, 
are comparable to those reported by other authors.  For instance, 
\citet{HSW84_2}
saw relative errors of 1-3\% in density and velocity near the horizon
using an artificial viscosity code on a $25 \times 10$ cylindrical grid.
\citet{Banyuls97} saw 
mean relative errors of 2.67\% and 0.99\% using a Godunov-type 
method on a $200 \times 5$ spherical grid.
Also, decreasing the Courant factor from $k_c=0.6$ 
to 0.2 reduces the errors in both
AV and NOCD methods by about a factor of three, consistent with
first order time discretization, and increases the rate of spatial
convergence closer to second order.

\section{Conclusion}
\label{sec:conclusion}

We have developed new artificial viscosity and
non-oscillatory central difference numerical
hydrodynamics schemes as integral components of the 
Cosmos code framework for performing fully general
relativistic calculations of strong field flows.
These methods have been discussed at length here and
compared also with published state-of-the-art Godunov methods
on their abilities to model shock tube, wall shock
and black hole accretion problems.
We find that for shock tube problems at moderate to high boost factors, 
with velocities up to $V \sim 0.99$ and limited only by grid
resolution, internal energy formulations using
artificial viscosity methods compare quite
favorably with total energy schemes such as the
NOCD methods, the Godunov methods using the
Marquina, Roe, or Flux-split approximate Riemann solvers, and the piecewise
parabolic method with an exact Riemann solver.
However, AV methods can be somewhat sensitive to parameters
(e.g., viscosity coefficients, Courant factor, etc.)
and generally suspect at high boost factors ($V > 0.95$) in
the wall shock problems we have considered here. 
On the other hand, NOCD methods
can easily be extended to ultra-relativistic velocities ($1-V < 10^{-11}$)
for the same wall shock tests,
and are comparable in accuracy over the entire range of velocities
we have simulated to the more standard but complicated
Riemann solver codes. NOCD schemes thus provide a robust new alternative
to simulating relativistic hydrodynamical flows since they offer
the same advantages of Godunov
methods in capturing ultra-relativistic flows but without the
cost and complication of Riemann solvers or flux splitting. 
They also provide all the advantages of AV methods in their speed, ease of
implementation, and general applicability (including straightforward
extensions to more general equations of state)
without explicitly using artificial viscosity for shock capturing.

\begin{acknowledgements}
This work was performed
under the auspices of the U.S. Department of Energy by
University of California, Lawrence
Livermore National Laboratory under Contract W-7405-Eng-48.
\end{acknowledgements}

\clearpage

%%%%%%%%%%%%%%%%%%%%%% FIGURES %%%%%%%%%%%%%%%%%

% Fig 1 (stube_QFC.eps)

% Fig 2 (stube_HRSC.eps)

% Fig 3 (stube_QFC_boost.eps)

% Fig 4 (stube_HRSC_boost.eps)

% Fig 5 (stube_all.eps)

% Fig 6 (wshock_QFC.eps)

% Fig 7 (wshock_HRSC.eps)

% Fig 8 (wshock_all.eps)

% Fig 9 (wshock_HRSC_boost.eps)

% Fig 10 (dust_QFC-2.eps)

% Fig 11 (dust_HRSC-2.eps)

%%%%%%%%%%%%%%%%%%%% TABLES %%%%%%%%%%%%%%%%%%%%%
\clearpage
\begin{table}[b]
%\begin{minipage}[b]{6.5in}
\vskip30pt
\begin{tabular}{ccccc} \hline\hline
Grid  & Method  & $\Vert E(\rho) \Vert_1$  & $\Vert E(P) \Vert_1$  & $\Vert E(V) \Vert_1$  \\
  \hline
200      & AV      & $9.08\times 10^{-2}$   & $5.62\times 10^{-2}$   & $1.18\times 10^{-2}$  \\
         & NOCD    & $1.06\times 10^{-1}$   & $5.50\times 10^{-2}$   & $1.17\times 10^{-2}$  \\
         & Marquina\tablenotemark{a} 
                   & $7.65\times 10^{-2}$   & $4.60\times 10^{-2}$   & $8.13\times 10^{-3}$  \\
  \hline
400      & AV      & $4.90\times 10^{-2}$   & $3.00\times 10^{-2}$   & $5.59\times 10^{-3}$  \\
         & NOCD    & $4.60\times 10^{-2}$   & $2.00\times 10^{-2}$   & $4.13\times 10^{-3}$  \\
         & Marquina\tablenotemark{a} & $4.65\times 10^{-2}$      
                   & $2.41\times 10^{-2}$   & $4.84\times 10^{-3}$  \\
  \hline
800      & AV      & $3.23\times 10^{-2}$   & $1.86\times 10^{-2}$   & $3.74\times 10^{-3}$  \\
         & NOCD    & $2.97\times 10^{-2}$   & $1.35\times 10^{-2}$   & $2.67\times 10^{-3}$  \\
  \hline
$128^3$  & AV      & $1.43\times 10^{-1}$   & $1.43\times 10^{-1}$   & $7.40\times 10^{-3}$  \\
         & NOCD    & $6.36\times 10^{-2}$   & $6.74\times 10^{-2}$   & $4.84\times 10^{-3}$  \\
         & Marquina\tablenotemark{a}
                   & $9.23\times 10^{-2}$   & $7.98\times 10^{-2}$   & $9.66\times 10^{-3}$  \\
\hline\hline
\end{tabular}
\caption{$L$-1 norm errors in density, pressure, and velocity 
for the moderate boost shock-tube tests. }
\label{tab:errors1}
\tablenotetext{a}{\citet{Font00}}
%\end{minipage}
\end{table}

%\clearpage
\begin{table}[b]
%\begin{minipage}[b]{6.5in}
\vskip30pt
\begin{tabular}{cccccccc} \hline\hline
Grid  & Method  & $\Vert E(\rho) \Vert_1$  & $\Vert E(D) \Vert_1$  & $\Vert E(P) \Vert_1$ & $\Vert E(\tau) \Vert_1$  & $\Vert E(V) \Vert_1$  & $\Vert E(S) \Vert_1$  \\
  \hline
400      & AV      & $1.32\times 10^{-1}$   & & $2.35\times 10^0$   & & $1.18\times 10^{-2}$ &  \\
         & NOCD    & $1.70\times 10^{-1}$   & & $4.22\times 10^0$   & & $2.96\times 10^{-2}$ &  \\
         & PPM\tablenotemark{a}    
		   & & $3.21 \times 10^{-1}$  & & $4.10 \times 10^0$ & & $4.25 \times 10^{0}$ \\
  \hline
800      & AV      & $8.84\times 10^{-2}$   & & $1.48\times 10^0$   & & $7.07\times 10^{-3}$ &  \\
         & NOCD    & $1.30\times 10^{-1}$   & & $1.91\times 10^0$   & & $1.58\times 10^{-2}$ &  \\
         & PPM\tablenotemark{a}   
		   & & $1.78 \times 10^{-1}$ & & $2.67 \times 10^0$ & & $2.71 \times 10^0$ \\
  \hline
1600     & AV      & $5.94\times 10^{-2}$   & & $1.00\times 10^0$   & & $4.30\times 10^{-3}$ &  \\
         & NOCD    & $8.81\times 10^{-2}$   & & $1.10\times 10^0$   & & $8.06\times 10^{-3}$ &  \\
         & PPM\tablenotemark{a}   
		   & & $1.00 \times 10^{-1}$ & & $1.89 \times 10^0$ & & $1.83 \times 10^0$ \\
\hline\hline
\end{tabular}
\caption{$L$-1 norm errors in density, pressure, and velocity for 
the high boost shock-tube tests.  Quoted errors for the PPM method
represent the conserved quantities, not primitive variables.
}
\label{tab:errors2}
\tablenotetext{a}{\citet{Marti96}}
%\end{minipage}
\end{table}

%\clearpage
\begin{table}[b]
%\begin{minipage}[b]{6.5in}
\vskip30pt
\begin{tabular}{cccccc} \hline\hline
$P_L$  & Boost  & Method  & $\bar{\epsilon}_\mathrm{rel}(\rho)$  
       & $\bar{\epsilon}_\mathrm{rel}(P)$  & $\bar{\epsilon}_\mathrm{rel}(V)$  \\
  \hline
1.33   & 1.08   & AV     & $5.23\times 10^{-3}$  & $3.02\times 10^{-3}$  & $2.27\times 10^{-2}$  \\
       &        & NOCD   & $2.55\times 10^{-3}$  & $1.68\times 10^{-3}$  & $7.29\times 10^{-3}$  \\
  \hline
6.67   & 1.28   & AV     & $5.81\times 10^{-3}$  & $3.48\times 10^{-3}$  & $1.20\times 10^{-2}$  \\
       &        & NOCD   & $4.31\times 10^{-3}$  & $2.49\times 10^{-3}$  & $6.06\times 10^{-3}$  \\
  \hline
13.3   & 1.43   & AV     & $6.75\times 10^{-3}$  & $4.00\times 10^{-3}$  & $1.13\times 10^{-2}$  \\
       &        & NOCD   & $6.55\times 10^{-3}$  & $3.25\times 10^{-3}$  & $8.50\times 10^{-3}$  \\
  \hline
26.7   & 1.63   & AV     & $7.10\times 10^{-3}$  & $3.96\times 10^{-3}$  & $8.51\times 10^{-3}$  \\
       &        & NOCD   & $8.75\times 10^{-3}$  & $3.57\times 10^{-3}$  & $9.06\times 10^{-3}$  \\
  \hline
66.7   & 1.96   & AV     & $1.12\times 10^{-2}$  & $5.23\times 10^{-3}$  & $1.05\times 10^{-2}$  \\
       &        & NOCD   & $1.36\times 10^{-2}$  & $4.27\times 10^{-3}$  & $1.27\times 10^{-2}$  \\
  \hline
133.3  & 2.28   & AV     & $1.43\times 10^{-2}$  & $5.76\times 10^{-3}$  & $1.14\times 10^{-2}$  \\
       &        & NOCD   & $1.93\times 10^{-2}$  & $4.70\times 10^{-3}$  & $1.60\times 10^{-2}$  \\
  \hline
266.7  & 2.66   & AV     & $1.64\times 10^{-2}$  & $5.66\times 10^{-3}$  & $1.12\times 10^{-2}$  \\
       &        & NOCD   & $2.57\times 10^{-2}$  & $5.07\times 10^{-3}$  & $1.92\times 10^{-2}$  \\
  \hline
666.7  & 3.28   & AV     & $2.23\times 10^{-2}$  & $5.81\times 10^{-3}$  & $1.13\times 10^{-2}$  \\
       &        & NOCD   & $3.48\times 10^{-2}$  & $5.91\times 10^{-3}$  & $2.48\times 10^{-2}$  \\
  \hline
1333.3 & 3.85   & AV     & $2.66\times 10^{-2}$  & $6.33\times 10^{-3}$  & $1.31\times 10^{-2}$  \\
       &        & NOCD   & $4.21\times 10^{-2}$  & $6.82\times 10^{-3}$  & $3.04\times 10^{-2}$  \\
  \hline
2666.7 & 4.53   & AV     & $3.22\times 10^{-2}$  & $5.72\times 10^{-3}$  & $1.23\times 10^{-2}$  \\
       &        & NOCD   & $4.81\times 10^{-2}$  & $7.85\times 10^{-3}$  & $3.65\times 10^{-2}$  \\
  \hline
6666.7 & 5.63   & AV     & $3.66\times 10^{-2}$  & $5.99\times 10^{-3}$  & $1.21\times 10^{-2}$  \\
       &        & NOCD   & $5.22\times 10^{-2}$  & $9.35\times 10^{-3}$  & $4.57\times 10^{-2}$  \\
\hline\hline
\end{tabular}
\caption{Mean-relative errors in the primitive variables for 
different boost factors in the shock-tube test using an 800 zone grid. }
\label{tab:errors3}
%\end{minipage}
\end{table}

%\clearpage
\begin{table}[b]
\vskip30pt
\begin{tabular}{ccccc} \hline\hline
Grid  & Method  & $\Vert E(\rho) \Vert_1$  & $\Vert E(P) \Vert_1$  & $\Vert E(V) \Vert_1$  \\
  \hline
200      & AV   & $1.56(0.57)\times 10^{-1}$  & $3.23\times 10^{-2}$  & $5.48\times 10^{-3}$  \\
         & NOCD & $5.10(1.08)\times 10^{-2}$  & $1.60\times 10^{-2}$  & $3.34\times 10^{-3}$  \\
  \hline
400      & AV   & $1.06(0.29)\times 10^{-1}$  & $2.18\times 10^{-2}$  & $2.59\times 10^{-3}$  \\
         & NOCD & $3.26(0.48)\times 10^{-2}$  & $1.10\times 10^{-2}$  & $2.69\times 10^{-3}$  \\
  \hline
800     & AV    & $9.51(1.43)\times 10^{-2}$  & $2.40\times 10^{-2}$  & $2.07\times 10^{-3}$  \\
        & NOCD  & $1.74(0.22)\times 10^{-2}$  & $6.26\times 10^{-3}$  & $1.50\times 10^{-3}$  \\
\hline\hline
\end{tabular}
\caption{$L$-1 norm errors for the relativistic
wall shock test with infall velocity $V=-0.9c$.  The values given in parentheses are the 
contribution of the first 20 zones to the total error.  
Wall heating dominates and greatly inflates the errors in regions
near the reflective boundary, especially in the AV methods.
}
\label{tab:errors4}
\end{table}

%\clearpage
\begin{table}[b]
%\begin{minipage}[b]{6.5in}
\vskip30pt
\begin{tabular}{ccccc} \hline\hline
$\nu$   & Method  & $\bar{\epsilon}_\mathrm{rel}(\rho)$  
                 & $\bar{\epsilon}_\mathrm{rel}(P)$  & $\bar{\epsilon}_\mathrm{rel}(V)$  \\
  \hline
0.4   & AV             & $5.40\times 10^{-2}$  & $4.71\times 10^{-2}$  & $6.33\times 10^{-3}$  \\
      & NOCD           & $2.07\times 10^{-2}$  & $2.48\times 10^{-2}$  & $9.91\times 10^{-3}$  \\
      & Wilson\tablenotemark{a}
                       & $5.36\times 10^{-2}$  &      &   \\
  \hline
0.17  & AV             & $8.09\times 10^{-2}$  & $6.42\times 10^{-2}$  & $2.14\times 10^{-2}$  \\
      & NOCD           & $1.27\times 10^{-2}$  & $1.15\times 10^{-2}$  & $9.31\times 10^{-3}$  \\
      & Wilson\tablenotemark{a}
                       & $6.98\times 10^{-2}$  &      &   \\
  \hline
0.1   & AV             & $9.59\times 10^{-2}$  & $7.44\times 10^{-2}$  & $3.66\times 10^{-2}$  \\
      & NOCD           & $8.95\times 10^{-3}$  & $7.23\times 10^{-3}$  & $6.41\times 10^{-3}$  \\
      & Wilson\tablenotemark{a}
                       & $8.29\times 10^{-2}$  &      &   \\
      & Marquina\tablenotemark{b}    
                       & $9.66\times 10^{-3}$  & $9.07\times 10^{-3}$  & $8.03\times 10^{-3}$  \\
  \hline
0.05  & AV             & $1.16\times 10^{-1}$  & $8.51\times 10^{-2}$  & $5.76\times 10^{-2}$  \\
      & NOCD           & $7.69\times 10^{-3}$  & $6.12\times 10^{-3}$  & $6.74\times 10^{-3}$  \\
  \hline
0.03  & AV             & $1.33\times 10^{-1}$  & $9.38\times 10^{-2}$  & $7.38\times 10^{-2}$  \\
      & NOCD           & $9.40\times 10^{-3}$  & $7.25\times 10^{-3}$  & $1.01\times 10^{-2}$  \\
  \hline
$10^{-3}$   & NOCD         & $4.43\times 10^{-3}$  & $2.73\times 10^{-3}$  & $4.60\times 10^{-3}$  \\
       & Marquina\tablenotemark{b}
                       & $7.20\times 10^{-3}$  & $5.80\times 10^{-3}$  & $1.26\times 10^{-2}$  \\
  \hline
$10^{-5}$   & NOCD       & $2.09\times 10^{-3}$  & $1.01\times 10^{-3}$  & $1.35\times 10^{-3}$  \\
       & Marquina\tablenotemark{b}
                       & $7.93\times 10^{-3}$  & $1.00\times 10^{-3}$  & $7.20\times 10^{-3}$  \\
  \hline
$10^{-7}$   & NOCD     & $6.30\times 10^{-3}$  & $5.59\times 10^{-3}$  & $1.29\times 10^{-2}$  \\
       & Marquina\tablenotemark{b}
                       & $9.30\times 10^{-3}$  & $6.10\times 10^{-3}$  & $8.56\times 10^{-3}$  \\
  \hline
$10^{-9}$   & NOCD   & $5.82\times 10^{-3}$  & $5.14\times 10^{-3}$  & $9.97\times 10^{-3}$  \\
       & Marquina\tablenotemark{b}
                       & $1.03\times 10^{-2}$  & $6.52\times 10^{-3}$  & $8.13\times 10^{-3}$  \\
  \hline
$10^{-11}$  & NOCD  & $1.12\times 10^{-3}$  & $8.27\times 10^{-4}$  & $5.08\times 10^{-4}$  \\
       & Marquina\tablenotemark{b}
                       & $3.40\times 10^{-2}$  & $1.41\times 10^{-3}$  & $3.26\times 10^{-3}$  \\
\hline\hline
\end{tabular}
\caption{Mean-relative errors in density, pressure, and velocity over a 
broad range of
infall velocities ($\vert V \vert = 1-\nu$) in the wall shock test using a 
200 zone grid. 
As noted in the text, the AV errors can be reduced significantly and brought
closer in agreement with the NOCD results by either increasing the
viscosity strength or decreasing the Courant factor.
}
\label{tab:errors5}
\tablenotetext{a}{\citet{CW84}}
\tablenotetext{b}{\citet{Aloy99}}
%\end{minipage}
\end{table}

%\clearpage
\begin{table}[b]
%\begin{minipage}[b]{6.5in}
\vskip30pt
\begin{tabular}{cccc} \hline\hline
Grid   & Method  & $\bar{\epsilon}_\mathrm{rel}(\rho)$  
       & $\bar{\epsilon}_\mathrm{rel}(V)$  \\
  \hline
$16^3$ & AV      & $4.81\times 10^{-2}$  & $3.08\times 10^{-2}$  \\
       & NOCD    & $1.01\times 10^{-1}$  & $1.53\times 10^{-2}$  \\
  \hline
$32^3$ & AV      & $2.70\times 10^{-2}$  & $1.34\times 10^{-2}$  \\
       & NOCD    & $4.55\times 10^{-2}$  & $3.26\times 10^{-3}$  \\
  \hline
$64^3$ & AV      & $1.36\times 10^{-2}$  & $6.32\times 10^{-3}$  \\
       & NOCD    & $2.11\times 10^{-2}$  & $1.44\times 10^{-3}$  \\
\hline\hline
\end{tabular}
\caption{Mean-relative errors in density and velocity for 
the black hole accretion problem at time $t=50M$, where $M=1$
is the black hole mass. }
\label{tab:errors6}
%\end{minipage}
\end{table}


\clearpage
\begin{thebibliography}{10}


\bibitem[Aloy,  Ib\'a\~nez, \& Mart\'i (1999)]{Aloy99}
Aloy, M. A.,  Ib\'a\~nez, J. M., and Mart\'i, J. M. 1999, \apjs, 
122, 151

\bibitem[Anninos (1998)]{Anninos98}
Anninos, P. 1998, \prd, 58, 064010

\bibitem[Banyuls et al. (1997)]{Banyuls97}
Banyuls, F., Font, J. A., Ib\'a\~nez, J. M., Mart\'i, J. M. and Miralles, 
J. A. 1997, \apj, 476, 221

\bibitem[Centrella \& Wilson (1984)]{CW84}
Centrella, J. and Wilson, J. R. 1984, \apjs, 54, 229

\bibitem[Colella \& Woodward (1984)]{Colella84}
Colella, P. and Woodward, P. R. 1984, J. Comput. Phys., 54, 174

\bibitem[Donat \& Marquina (1996)]{Donat96}
Donat, R. and Marquina, A. 1996, J. Comput. Phys., 125, 42

\bibitem[Eulderink \& Mellema (1995)]{Eulderink95}
Eulderink, F. and Mellema, G. 
1995, Astron. Astrophys. Suppl. Ser., 110, 587

\bibitem[Font et al. (2000)]{Font00}
Font, J. A., Miller, M., Suen, W. and Tobias, M. 2000, \prd, 61,
 044011

\bibitem[Hawley, Smarr, \& Wilson (1984a)]{HSW84_1}
Hawley, J. F., Smarr, L. L. and Wilson, J. R. 1984, \apj, 277, 296

\bibitem[Hawley, Smarr, \& Wilson (1984b)]{HSW84_2}
Hawley, J. F., Smarr, L. L. and Wilson, J. R. 1984, \apjs, 55, 211

\bibitem[Jiang et al. (1998)]{Jiang98a}
Jiang, O.-S., Levy, D., Lin, C.-T., Osher, S. and Tadmor, E.
1998, SIAM J. Numer. Anal., 35, 2147

\bibitem[Jiang \& Tadmor (1998)]{Jiang98b}
Jiang, O.-S. and Tadmor, E.
1998, SIAM J. Sci. Comput., 19, 1892

\bibitem[Mart\'i \& M\"uller (1996)]{Marti96}
Mart\'i, J. and M\"uller, E. 1996, J. Comput. Phys., 123, 1

\bibitem[May \& White (1966)]{May66}
May, M. M. and White, R. H. 1966, \prd, 141, 1232

\bibitem[May \& White (1967)]{May67}
May, M. M. and White, R. H. 1967, Meth. Computat. Phys., 7, 219

\bibitem[Norman \& Winkler (1986)]{Norman86}
Norman, M. L. and Winkler, K.-H. A. 1986,
in Astrophysical Radiation Hydrodynamics, ed. M. L. Norman \& K.-H. A. Winkler
(NATO ASI ser. C, 188) (Dordrecht:Reidel), 449

\bibitem[Quirk (1994)]{Quirk94}
Quirk, J. J., 1994, Int. J. for Numerical 
Methods in Fluids, 18, 555

\bibitem[Thompson (1986)]{Thompson86}
Thompson, K. W. 1986, J. Fluid Mech., 171, 365

\bibitem[Tscharnuter \& Winkler (1979)]{TW79}
Tscharnuter, W.-M. and Winkler, K.-H. 
1979, Comput. Phys. Comm., 18, 171

\bibitem[van Leer (1977)]{VL77}
van Leer, B. 1977, J. Comput. Phys., 23, 276

\bibitem[VonNeumann \& Richtmyer (1950)]{Neumann50}
VonNeumann, J. and Richtmyer, R. D. 1950, J. Appl. Phys., 21, 232

\bibitem[Wilson (1972)]{Wilson72}
Wilson, J. R. 1972, \apj, 173, 431

\bibitem[Wilson (1979)]{Wilson79}
Wilson, J. R., in {\it Sources of Gravitational Radiation},
edited by L. Smarr (Cambridge University Press, Cambridge, England, 1979)

\end{thebibliography}
\end{document}